\documentclass[acmsmall]{acmart}

\AtBeginDocument{%
  \providecommand\BibTeX{{%
    \normalfont B\kern-0.5em{\scshape i\kern-0.25em b}\kern-0.8em\TeX}}}

\usepackage{natbib}
\usepackage{moresize}
\usepackage{amsmath}
\usepackage[latin1]{inputenc}
\usepackage{verbatim}
\usepackage{multirow}
\usepackage{subcaption}
\usepackage[normalem]{ulem}
\usepackage{array}
\usepackage{listings}
\usepackage{wrapfig}
\usepackage[labelfont=bf]{caption}

\usepackage[nodisplayskipstretch]{setspace}

\usepackage{enumitem}

\newcommand{\textcodeit}[1]{\textit{\ttfamily\small #1}}

\newcommand{\textcode}[1]{\texttt{\ttfamily\small #1}}

\usepackage{amsthm} 

\begin{document}

 \title{Statically Detecting Vulnerabilities by Processing Programming Languages as Natural Languages}

\author{Ib\'{e}ria Medeiros}
\affiliation{\institution{LASIGE, Faculdade de Ci\^{e}ncias, Universidade de Lisboa - Portugal}}
\email{imedeiros@di.fc.ul.pt}

\author{Nuno Neves}
\affiliation{\institution{LASIGE, Faculdade de Ci\^{e}ncias, Universidade de Lisboa - Portugal}}
\email{nuno@di.fc.ul.pt}

\author{Miguel Correia}
\affiliation{\institution{INESC-ID, Instituto Superior T\'{e}cnico, Universidade de Lisboa - Portugal}}
\email{miguel.p.correia@tecnico.ulisboa.pt}

\renewcommand{\shortauthors}{Medeiros, et al.}

\begin{abstract}
Web applications continue to be a favorite target for hackers due to a combination of wide adoption and rapid deployment cycles, which often lead to the introduction of high impact vulnerabilities.
Static analysis tools are important to search for bugs automatically in the program source code, supporting developers on their removal.
However, building these tools requires programming the knowledge on how to discover the vulnerabilities.
This paper presents an alternative approach in which tools \emph{learn} to detect flaws automatically by resorting to artificial intelligence concepts, more concretely to natural language processing. 
The approach employs a sequence model to learn to characterize vulnerabilities based on an annotated corpus.
Afterwards, the model is utilized to discover and identify vulnerabilities in the source code.
It was implemented in the DEKANT tool and evaluated experimentally with a large set of PHP applications and WordPress plugins. Overall, we found several hundred vulnerabilities belonging to 12 classes of input validation vulnerabilities, where 62 of them were zero-day.
\end{abstract}

\ccsdesc[500]{Security and privacy~Vulnerability management}
\ccsdesc[500]{Security and privacy~Web application security}
\ccsdesc[500]{Computing methodologies~Natural language processing}
\ccsdesc[500]{Computing methodologies~Speech recognition}

\maketitle

\section{Introduction}
\label{sec:introduction}

Web applications are being used to implement interfaces of a myriad of services. They are often the first target of attacks, and despite considerable efforts to improve security, there are still many examples of high impact  compromises.
In the 2017 OWASP Top 10 list, vulnerabilities like SQL injection (SQLI) and cross-site scripting (XSS) continue to raise significant concerns, but other classes are also listed as being commonly exploited~\cite{owasp:17}. Millions of websites have been compromised since Oct.~2014 due to vulnerabilities in plugins of Drupal~\cite{BBC:14} and WordPress~\cite{HackerNews:17,  ThreatPost:17},  and the data of more than a billion users has been stolen using SQLI attacks against various kinds of services (governmental, financial, education, mail, etc) \cite{HackerNews:17a, HelpNetSecurity:17}. In addition, the next wave of XSS attacks has been predicted for the past two years, with an important expected growth of the problem~\cite{Sink:17,Imperva:17}.

Many of these vulnerabilities are related to malformed inputs that reach some relevant asset (e.g., the database or the user's browser) by traveling through a  code \emph{slice} (a series of instructions) of the web application. Therefore, a good practice to enhance security is to pass inputs through \emph{sanitization functions} that invalidate dangerous metacharacters or/and \emph{validation functions} that check their content.
In addition, programmers commonly use \emph{static analysis tools} to search automatically for bugs in the source code, facilitating their removal. The development of these tools, however, requires coding explicitly  the knowledge on how each vulnerability can be detected \cite{Dahse:14, Fonseca:14, Jovanovic:06, Medeiros:15a}, which is a complex task. Moreover, this knowledge might be incomplete or partially wrong, making the tools inaccurate \cite{Dahse:15}. For example, if the tools do not understand that a certain function sanitizes inputs, they could raise an alert about a vulnerability that does not exist.

This paper presents a new approach for static analysis that is based on \emph{learning to recognize vulnerabilities}. It leverages from artificial intelligence (AI) concepts, more precisely from classification models for sequences of observations that are commonly used in the field of natural language processing (NLP). 
NLP is a confluence of AI and linguistics, which involves intelligent analysis of written language, i.e., the natural languages. In this sense, NLP is considered a sub-area of AI. It can be viewed as a new form of intelligence in an artificial way that can get insights how humans understand natural languages.
NLP tasks, such as parts-of-speech (PoS) tagging or named entity recognition (NER), are typically modelled as sequence classification problems, in which a class (e.g., a given morpho-syntactic category) is assigned to each word in a given sentence, according to estimate given by a structured prediction model that takes word order into consideration. The model's parameters are normally inferred using supervised machine learning techniques, taking advantage of annotated corpora.

We propose applying a similar approach to web programming languages, i.e., to analyse source code in a similar manner to what is being done with natural language text. Even though, these languages are artificial, they have many characteristics in common with natural languages, such as words, syntactic rules,  sentences, and a grammar.
NLP usually employs machine learning to extract rules (knowledge) automatically from a \emph{corpus}.
Then, with this knowledge, other sequences of observations can be processed and classified. NLP has to take into account the \emph{order} of the observations, as the meaning of sentences depends on it. Therefore NLP involves forms of classification more sophisticated than approaches based on \emph{standard classifiers} (e.g., naive Bayes, decision trees, support vector machines), which simply check the presence of certain observations without considering any relation between them.

Our approach for static analysis resorts to machine language techniques that take the order of source code instructions into account -- \emph{sequence models} -- to allow accurate detection and identification of the vulnerabilities in the code.
Previous applications of machine learning in the context of static analysis neither produced tools that learn to make detection nor employed sequence models.
For example, PhpMinerII resorts to machine learning to train standard classifiers, which then verify if certain constructs (associated with flaws) exist in the code. However, it does not provide the exact location of the vulnerabilities \cite{Shar:12a,Shar:12b}.
WAP and WAPe use a taint analyser to search for vulnerabilities and a standard classifier to confirm that the found bugs\footnote{In software security context, we consider vulnerability as a being a bug or a flaw that can be exploitable.} can actually create security problems  \cite{Medeiros:15a}.
None of these tools considers the order of code elements or the relation among them, leading to bugs being missed (\emph{false negatives, FN}) and alarms being raised on correct code (\emph{false positives, FP}).

Our sequence model is a \emph{Hidden Markov Model} (HMM) \cite{Rabiner:89}. A HMM is a Bayesian network composed of nodes corresponding to the states and edges associated to the probabilities of transitioning  between states. States are hidden, i.e., are not observed. Given a sequence of observations, the hidden states (one per observation) are discovered following the model and taking into account the order of the observations. Therefore, the HMM can be used to find the series of states that \emph{best} explains the sequence of observations. 

The paper also presents the \emph{hidDEn marKov model diAgNosing vulnerabiliTies} (DEKANT) tool that implements our approach for applications written in PHP. 
The tool was evaluated experimentally with a diverse set of 23 open source web applications with bugs disclosed in the past. These applications are substantial, with an aggregated size of around 8,000 files and 2.5 million lines of code (LoC). All flaws that we are aware of being previously reported were found by DEKANT. More than one thousand slices were analyzed, 714 were classified as having vulnerabilities and 305 as not. The false positives were in the order of two dozens.
In addition, the tool checked 23 plugins of WordPress  and found \emph{62 zero-day vulnerabilities}. These flaws were reported to the developers, and some of them already confirmed their existence and fixed the plugins.  DEKANT was also assessed with several other vulnerability detection tools, and the results give evidence that our approach leads to better accuracy and precision.

The main contributions of the paper are:
(1) a novel approach for improving the security of web applications by letting static analysis tools learn to discover vulnerabilities through an annotated corpus;
(2) an intermediate language representation capturing the relevant features of  PHP, and a sequence model that takes into consideration the place where code elements appear in the slices and how they alter the spreading of the input data;
(3) a static analysis tool that implements the approach;
(4) an experimental evaluation that demonstrates the ability of this tool to find known and zero-day vulnerabilities with a residual number of mistakes.


\section{Related Work}
\label{s:related-work}

Static analysis tools search for vulnerabilities in the applications usually by processing the source code (e.g.,  \cite{Fonseca:14,Jovanovic:06,Shankar:01,Son:11,Dahse:14, Medeiros:15a,Backes:17}). 
Many of these tools perform taint analysis, tracking user inputs to determine if they reach a sensitive sink (i.e., a function that could be exploited).
Pixy \cite{Jovanovic:06} was one of the first tools to automate this kind of analysis on PHP applications. Later on, RIPS \cite{Dahse:14} extended this technique with the ability to process more advanced constructs of PHP (e.g., objects). phpSAFE \cite{Fonseca:14} is a recent solution that does taint analysis to look for flaws in CMS plugins (e.g., WordPress plugins).
WAP \cite{Medeiros:15a, Medeiros:16} also does taint analysis, but aims at reducing the number of false positives by resorting to  data mining, besides also correcting automatically the located bugs.
Other works \cite{Yamaguchi:14, Yamaguchi:15} detect vulnerabilities by processing source code properties represented as graphs.
In this paper, we propose an novel approach which, unlike these works, does not involve programming information about bugs, but instead extracts this knowledge from annotated code samples and thus learns to find the vulnerabilities. 

Machine learning has been used in a few works to measure the quality of software by collecting a series of attributes that reveal the presence of software defects \cite{Arisholm:10,Lessmann:08}. 
Other approaches resort to machine learning to predict if there are vulnerabilities in a program~\cite{Neuhaus:07,Walden:09, Perl:15},
which is different from identifying precisely the bugs, something that we do in this paper. To support the predictions they employ various features, such as past vulnerabilities and function calls~\cite{Neuhaus:07},
or a combination of code-metric analysis with metadata gathered from application repositories~\cite{Perl:15}. In particular, PhpMinerI and PhpMinerII predict the presence of vulnerabilities in PHP programs \cite{Shar:12a,Shar:12b, Shar:13}. The tools are first trained with a set of annotated slices that end at a sensitive sink (but do not necessarily start at an entry point), and then they are ready to identify slices with errors. 
WAP and WAPe are different because they use machine learning and data mining to predict if a vulnerability detected by taint analysis is actually a real bug or a false alarm \cite{Medeiros:15a, Medeiros:16}. In any case, PhpMiner and WAP tools employ standard classifiers (e.g., Logistic Regression or a Multi-Layer Perceptron) instead of structured prediction models (i.e., a sequence classifier) as we propose here.

There are a few static analysis tools that implement machine learning techniques.
Chucky~\cite{Yamaguchi:13} discovers vulnerabilities by identifying missing checks in C language software. 
VulDeePecker~\cite{Li:18} resorts to code gadgets to represent parts of C programs and then transforms them into vectors. A neural network system then determines if the target program is vulnerable due to buffer or resource management errors. Russell et al.~\cite{Russell:18} developed a vulnerability detection tool for C and C++ based on features learning from a dataset and artificial neural network.
Scandariato et al.~\cite{Scandariato:14} performs text mining to predict vulnerable software components in Android applications.  SuSi~\cite{Rasthofer:14} employs machine learning to classify sources and sinks in the code of Android API.

This paper extends our previous work~\cite{Medeiros:16a}. Our approach extracts PHP slices, but contrary to the others it translates them into a tokenized language to be processed by a HMM. While tools in the literature collect attributes from a slice and classify them without considering ordering relations among statements, which is simplistic, DEKANT also does classification but takes into account the place in which code elements appear in the slice. Such form of classification assists on a more accurate and precise detection of bugs.


\begin{table*}[b]
	\centering
	\ssmall\ttfamily
	\addtolength{\tabcolsep}{-4.5pt}
	\begin{tabular}{l|l|l|l|l}
		\hline
		\hspace{-0.3cm} \textbf{PHP code}	&	\textbf{slice-isl} &	\textbf{variable map} &	\textbf{tainted list} & \textbf{slice-isl classification}\\ \hline 
		\hspace{-0.5cm} 1 \$u = \$\_POST[`username'];	&	input var 	&	1 - u 		&	TL = \{u\}	&	$\langle$input,Taint$\rangle$ $\langle$var\_vv\_u,Taint$\rangle$ \\
		\hspace{-0.5cm} 2 \$q = "SELECT pass FROM users WHERE user='".\$u."'";	&	var var		&	1 u q 		&	TL = \{u, q\}	&	$\langle$var\_vv\_u,Taint$\rangle$ $\langle$var\_vv\_q,Taint$\rangle$ \\
		\hspace{-0.5cm} 3 \$r = mysqli\_query(\$con, \$q);	&	ss var var	&	1 - q r	&	TL = \{u, q, r\}	&	$\langle$ss,N-Taint$\rangle$ $\langle$var\_vv\_q,Taint$\rangle$ $\langle$var\_vv\_r,Taint$\rangle$ \\
		\hline
		\multicolumn{1}{c}{\scriptsize \textrm{(a) code with SQLI vulnerability}} & \multicolumn{3}{c}{\scriptsize \textrm{(b) \emph{slice-isl}}} & \multicolumn{1}{c}{\scriptsize \textrm{(c) outputting the final classification}}\\
	\end{tabular}
	\vspace{-0.2cm}
	\captionof{figure}{\small Code vulnerable to SQLI, translation into ISL, and detection of the vulnerability.}
	\label{t:classif_sqli_1}
\end{table*}

\begin{table*}[b]
	\centering
	\ssmall\ttfamily
	\addtolength{\tabcolsep}{-4.5pt}
	\begin{tabular}{l|l|l|l}
		\hline
		\hspace{-0.3cm} \textbf{PHP code}	&	\textbf{slice-isl} &	\textbf{variable map} &	\textbf{list} \\ \hline 
		\hspace{-0.5cm} 1 \$u = (isset(\$\_POST[`name']) ? \$u = \$\_POST[`name'] : '';	&	input var 	&	1 - u 	&	TL = \{u\}; CTL = \{\} \\
		\hspace{-0.5cm} 2 \$a = \$\_POST[`age'];	&	input var 	&	1 - a	&	TL = \{u, a\}; CTL = \{\} \\
		\hspace{-0.5cm} 3 if (isset(\$a) \&\& preg\_match('/[a-zA-Z]+/', \$u) \&\&	&	cond fillchk var contentchk var		&	0 - - a - u - a -	&	TL = \{u, a\}; CTL = \{u, a\}	 \\
		\hspace{-0.5cm} \hspace{0.75cm} is\_int(\$a))	&	\hspace{0.67cm}typechk var cond		&	&		 \\	
		\hspace{-0.5cm} 4 \hspace{0.2cm} echo '$<$input type="hidden" name="user" value="'.\${u}.'"$>$';	&	cond ss var	&	0 - - u &	TL = \{u, a\}; CTL = \{u, a\}	 \\	
		\hspace{-0.5cm} 5 else &	cond	&	0 - &	TL = \{u, a\}; CTL = \{\}	 \\
		\hspace{-0.5cm} 6 \hspace{0.2cm} echo \${u} . "is an invalid user";	&	ss var	&	0 - u &	TL = \{u, a\}; CTL = \{\} \\
		\hline
		\multicolumn{1}{c}{\scriptsize\textrm{(a) code with XSS vulnerability and validation}} & \multicolumn{2}{c}{\scriptsize\textrm{(b) \emph{slice-isl} and variable map}} & \multicolumn{1}{c}{\scriptsize\textrm{(c) artefacts lists}}\\
	\end{tabular}
	
	\vspace{-0.2cm}
	\captionof{figure}{\small Code with a slice vulnerable to XSS (lines \{1, 3, 5, 6\}) and a slice not vulnerable  (lines \{1, 2, 3, 4\}), with ISL translation.}
	\label{t:xss}
	\vspace{-0.4cm}
\end{table*}

\section{Surface Vulnerabilities}
\label{prob:vvs}

Many classes of security flaws in web applications are caused by improper handling of user inputs. Therefore, they are denominated \emph{surface vulnerabilities} or \emph{input validation vulnerabilities}.  In PHP programs the malicious input arrives to the application (e.g, \emph{\$\_POST}), then it may suffer various modifications and might be copied to variables, and eventually reaches a security-sensitive function (e.g.,  \emph{mysqli\_query} or \emph{echo}) inducing an erroneous action. 
Below, we introduce the 12 classes of surface vulnerabilities that will be considered in rest of the paper. 

SQLI is the class of  vulnerabilities with highest risk in the OWASP Top 10 list \cite{owasp:17}. Normally, the  malicious input is used to change the behavior of a query to a database to provoke the disclosure of private data or corrupt the tables.

\begin{example}
	The PHP script of Fig. \ref{t:classif_sqli_1} (a) has a simple SQLI vulnerability.
	\textcode{\$u} receives the username provided by the user (line 1), and then it is inserted in a query (lines 2-3). An attacker can inject a malicious username like \textcode{' OR 1 = 1 -\,- }, 
	modifying the structure of the query and getting the passwords of all users.
\end{example}

XSS vulnerabilities allow attackers to execute scripts in the users' browsers. Below we give an example: 

\begin{example}
	The code snippet of Fig.~\ref{t:xss} (a) has a XSS vulnerability.
	If the user provides a name, it gets saved in \textcode{\${u}} (line 1). Then, if conditional validation is false (line 3), the value is returned to the user by \textcodeit{echo} (line 6). A script provided as input would be executed in the browser, possibly carrying out some malicious deed.
\end{example}

The other classes are presented briefly.
Remote and local file inclusion (RFI/LFI) flaws also allow attackers to insert code in the vulnerable web application. While in RFI the code can be located in another web site, in LFI it has to be in the local file system (but there are also several strategies to put it there). 
OS command injection (OSCI) lets an attacker to provide commands to be run 
in a shell of the OS of the web server. Attackers can supply code that is executed by a \textcodeit{eval} function by exploring PHP command injection (PHPCI) bugs. LDAP injection (LDAPI), like SQLI, is associated to the construction and execution of queries, in this case for the LDAP service. An attacker can read files from the local file system by exploiting directory traversal / path traversal (DT/PT) and source code disclosure (SCD) vulnerabilities. A comment spamming (CS) bug is related to the ranking manipulation of spammers' web sites. Header injection or HTTP response splitting (HI) allows an attacker to manipulate the HTTP response. An attacker can force a web client to use a session ID he defined by exploiting a session fixation (SF) flaw.


\section{Overview of the Approach}
\label{s:approach_v2}

\begin{figure}[b]
	\centering
	\includegraphics[width=0.8\columnwidth]{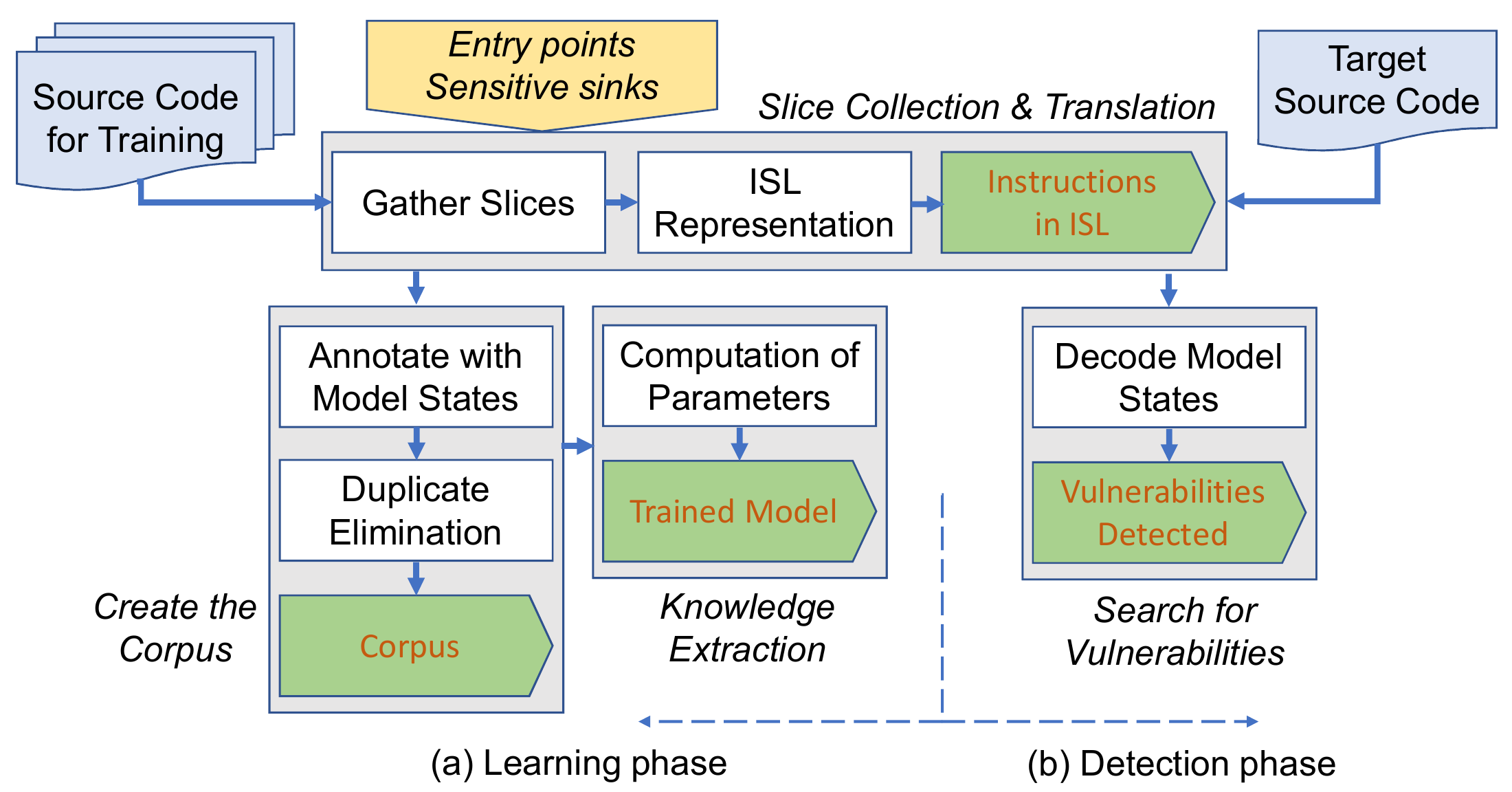}
	\vspace{-4mm}
	\caption{\small Overview on the proposed approach.}
	\label{f:approach_arch_2}
\end{figure}

Our approach for vulnerability detection examines program slices to determine if they contain a bug. The slices are collected from the source code of the target application, and then their instructions are represented in an intermediate language developed to express features that are relevant to surface vulnerabilities. Bugs are found by classifying the translated instructions with an HMM sequence model. Since the model has an understanding of how the data flows are affected by operations related to sanitization, validation and modification, it becomes feasible to make an accurate analysis. In order to setup the model, there is a learning phase where an annotated corpus is employed to derive the knowledge about the different classes of vulnerabilities. 
Afterwards, the model is used to detect vulnerabilities.
Fig.~\ref{f:approach_arch_2} illustrates this procedure. 

In more detail, the following steps are carried out. The \emph{learning phase} is composed mainly of steps (1)-(3) while the \emph{detection phase} encompasses (1) and (4):

(1) \emph{Slice collection and translation:} get the slices from the application source code (either for learning or detection). Since we are focusing on surface vulnerabilities, the only slices that have to be considered need to start at some point in the program where an user input is received (i.e., at an \emph{entry point}) and then they have to end at a security-sensitive instruction (i.e., a \emph{sensitive sink}). The resulting slice is a series of tracked instructions between the two points. Then, each instruction of a slice is represented into the \emph{Intermediate Slice Language} (ISL) (Section~\ref{s:ISL}). ISL is a categorized language with grammar rules that aggregate in classes the code elements by functionality. A slice in the ISL format is going to be named as \emph{slice-isl};

(2) \emph{Create the corpus:} build a corpus with a group of instructions represented in the intermediate language, which are labeled either as vulnerable or non-vulnerable. The instructions are provided individually or gathered from slices of training programs. Overall, the corpus includes both representative pieces of programs that have various kinds of flaws and that handle inputs adequately;

(3) \emph{Knowledge extraction:} acquire knowledge from the corpus to configure the HMM sequence model, namely compute the probability matrices;

(4) \emph{Search for Vulnerabilities:}  use the model to find the best sequence of states that explains a slice in the intermediate language. Each instruction in the slice corresponds to a sequence of observations. These observations are classified by the model, tracking the variables from the previous instructions to find out which emission probabilities are selected.
The state computed for the last observation of the last instruction determines the overall classification, either as vulnerable or not. If a flaw is found, an alert is reported including the location in the source code.

The next two sections explain the ISL language and the sequence model (Section~\ref{s:ISL} and \ref{s:hmm_ta}). Then, the four above steps are elaborated in Section~\ref{s:approach}. An overview of the tool that implements our approach is given in Section~\ref{s:impl_eval_dekant}.

\section{Intermediate Slice Language}
\label{s:ISL}

All slices commence with an entry point and finish with a sensitive sink; between them there can be an arbitrary number of statements, such as assignments that transmit data to intermediate variables and various kinds of expressions that validate or modify the data. In other words, a slice contains all instructions (lines of code) that manipulate and propagate an input arriving at an entry point and until a sensitive sink is reached, but no other statements.

ISL expresses an instruction into a few tokens. The instructions are composed of \emph{code elements} that are categorized in \emph{classes} of related items (e.g., class \textcodeit{input} takes PHP entry points like \textcode{\$\_GET} and \textcode{\$\_POST}). Therefore, classes are the \emph{tokens} of the ISL language and these are organized together accordingly to a grammar. Next we give a more careful explanation of ISL assuming that the source code is programmed in the PHP language. However, the approach is generic and other languages could be considered.

\subsection{Tokens}
\label{s:stl-tokens}

ISL abstracts away aspects of the PHP language that are irrelevant to the discovery of surface vulnerabilities. Therefore, as a starting point to specify ISL, it was necessary to identify the essential tokens. To achieve this, we followed an iterative approach where we began with an initial group of tokens which were gradually refined. In every iteration, we examined various slices (vulnerable and not) to recognize the important code elements. We also looked at the PHP instructions that could manipulate entry points and be associated to bugs or prevent them (e.g., functions that replace characters in strings). In addition, for PHP functions, we studied cautiously their parameters to determine which of them are crucial for our analysis. In the end, we defined around twenty tokens that are sufficient to describe the instructions of a PHP program. 

\begin{example}
	Function \emph{mysqli\_query} and its parameters correspond to two tokens: \textcodeit{ss} for sensitive sink; and \textcodeit{var} for variable or \textcodeit{input} if the parameter receives data originating from an entry point. Although this function has three parameters (the last of them optional), notice that just one of them (the second) is essential to represent.
\end{example}

\begin{table}[htb]\scriptsize
	\caption{\small Intermediate Slice Language tokens.}
	\label{t:ttokens}
	\centering \vspace{-2mm}	
	\begin{tabular}{|l|*{2}{l|}c|}							
		\hline							
		\textbf{Token}	&	\textbf{Description}	&	\textbf{PHP Function}	&	\textbf{Taint}	\\
		\hline\hline							
		
		input	&	entry point	&	\$\_GET, \$\_POST, \$\_COOKIE, \$\_REQUEST	&	Yes	\\
		&	&	\$\_HTTP\_GET\_VARS, \$\_HTTP\_POST\_VARS	&	\\
		&	&	\$\_HTTP\_COOKIE\_VARS, \$\_HTTP\_REQUEST\_VARS	&	\\
		&	&	\$\_FILES, \$\_SERVERS	&	\\ 
		var	&	variable	&	--	&	No	\\ 
		sanit\_f	&	sanitization function	&	mysql\_escape\_string, mysql\_real\_escape\_string	&	No	\\
		&	&	mysqli\_escape\_string, mysqli\_real\_escape\_string	&	\\
		&	&	mysqli\_stmt\_bind\_param, mysqli::escape\_string	&	\\
		&	&	mysqli::real\_escape\_string, mysqli\_stmt::bind\_param	&	\\
		&	&		&	\\
		&	&	htmlentities, htmlspecialchars, strip\_tags, urlencode	&	\\ 
		ss	&	sensitive sink	&	mysql\_query, mysql\_unbuffered\_query, mysql\_db\_query	&	Yes	\\
		&		&	mysqli\_query, mysqli\_real\_query, mysqli\_master\_query	&	\\
		&		&	mysqli\_multi\_query, mysqli\_stmt\_execute, mysqli\_execute	&	\\
		&		&	mysqli::query, mysqli::multi\_query, mysqli::real\_query	&	\\
		&		&	mysqli\_stmt::execute	&	\\
		&		&	&	\\
		&		&	fopen, file\_get\_contents, file, copy, unlink, move\_uploaded\_file	&	\\
		&		&	imagecreatefromgd2, imagecreatefromgd2part, imagecreatefromgd	&	\\
		&		&	imagecreatefromgif, imagecreatefromjpeg, imagecreatefrompng &	\\
		&		&	imagecreatefromstring, imagecreatefromwbmp	&	\\
		&		&	imagecreatefromxbm, imagecreatefromxpm	&	\\
		&		&	require, require\_once, include, include\_once		&	\\
		&		&		&	\\
		&		&	readfile	&	\\
		&		&		&	\\
		&		&	passthru, system, shell\_exec, exec, pcntl\_exec, popen	&	\\
		&		&		&	\\
		&		&	echo, print, printf, die, error, exit	&	\\
		&		&	file\_put\_contents, file\_get\_contents	&	\\
		&		&		&	\\ 
		&		&	eval	&	\\ 
		typechk\_str	&	type checking string function	&	is\_string, ctype\_alpha, ctype\_alnum	&	Yes	\\ 
		typechk\_num	&	type checking numeric function	&	is\_int, is\_double, is\_float, is\_integer	&	No	\\
		&		&	is\_long, is\_numeric, is\_real, is\_scalar, ctype\_digit	&	\\ 
		contentchk	&	content checking function	&	preg\_match, preg\_match\_all, ereg, eregi	&	No	\\
		&		&	strnatcmp, strcmp, strncmp, strncasecmp, strcasecmp	&	\\ 
		fillchk	&	fill checking function	&	isset, empty, is\_null	&	Yes	\\ 
		cond	&	\emph{if} instruction presence	&	if	&	No	\\ 
		join\_str	&	join string function	&	implode, join	&	No	\\ 
		erase\_str	&	erase string function	&	trim, ltrim, rtrim	&	Yes	\\ 
		replace\_str	&	replace string function	&	preg\_replace, preg\_filter, str\_ireplace, str\_replace	&	No	\\
		&		&	ereg\_replace, eregi\_replace, str\_shuffle, chunk\_split	&	\\ 
		split\_str	&	split string function	&	str\_split, preg\_split, explode, split, spliti	&	Yes	\\ 
		add\_str	&	add string function	&	str\_pad	&	Yes/No	\\ 
		sub\_str	&	substring function	&	substr	&	Yes/No	\\ 
		sub\_str\_replace	&	replace substring function	&	substr\_replace	&	Yes/No	\\ 
		char5	&	substring with less than 6 chars	&	--	&	No	\\ 
		char6	&	substring with more than 5 chars	&	--	&	Yes	\\ 
		start\_where	&	where the substring starts	&	--	&	Yes/No	\\ 
		conc	&	concatenation operator	&	--	&	Yes/No \\ \cline{1-4}
		var\_vv	&	variable tainted	&	--	&	Yes	\\
		miss	&	miss value	&	--	&	Yes/No	\\
		
		\hline							
	\end{tabular}
\end{table}

Table \ref{t:ttokens} summarizes the currently defined ISL tokens. The first column shows above the twenty tokens that stand for PHP code elements whereas the last two tokens are necessary only for the description of the corpus and the implementation of the model. The next two columns explain succinctly the purpose of the token and give a few examples. 
 Column four defines the taintedness status of each token which is used when building the corpus or performing the analysis.

A more cautious inspection of the tokens shows that they enable many relevant behaviors to be expressed. For example: Since the manipulation of strings plays a fundamental role in the exploitation of surface vulnerabilities, there are various tokens that enable a precise modeling of these operations (e.g., \textcodeit{erase\_str} or \textcodeit{sub\_str});
Tokens \textcodeit{char5} and \textcodeit{char6} act as the amount of characters that are manipulated by functions that extract or replace the contents from a user input; 
The place in a string  where modifications are applied (begin, middle or end) is described by \textcodeit{start\_where}; Token
\textcodeit{cond}  can correspond to an \textcodeit{if} statement that might have validation functions over variables (e.g., user inputs) as part of its conditional expression. This token allows the correlation among the validated variables and the variables that appear inside the \textcodeit{if} branches.

There are a few tokens that are \emph{context-sensitive}, i.e., whose selection depends not only on the code elements being translated but also on how they are utilized in the program.
Tokens \textcodeit{char5} and \textcodeit{char6} are two  examples as they depend on the substring length. 
If this length is only defined  at runtime, it is impossible to know precisely which token should be assigned. 
This ambiguity may originate errors in the analysis, either leading to false positives or false negatives. However, since we prefer to be conservative (i.e., report false positives instead of missing vulnerabilities), in the situation where the length is undefined, ISL uses the \textcodeit{char6} token because it allows larger payloads to be manipulated. Something similar occurs with the \textcodeit{contentchk} token that depends on the verification pattern.

ISL must be able to represent PHP instructions in all steps of the two phases of the approach. When slices are extracted for analysis, ISL sets all variables to the default token value \textcodeit{var}. However, when  instructions are placed in the corpus or are processed by the detection procedure, it is necessary to keep information about taintedness. In this case,  tainted and untainted variables are depicted respectively by the tokens \textcodeit{var\_vv} and \textcodeit{var}. The \textcodeit{miss} token is also used with the corpus and it serves to normalize the length of sequences (Section \ref{s:impl_eval_dekant}).

\subsection{Grammar}

The ISL grammar is specified by the rules in Listing \ref{f:stl_rules}. It allows the representation of the code elements included in the instructions into the tokens (Table \ref{t:ttokens}, column 3 entries are transformed into the column 1 tokens). A slice translated into ISL consists of a set of \textcodeit{statements} (line 2), each one defined by either: a rule that covers various operations like string concatenation (lines 4-11); or an conditional (line 12); or an assignment (line 13). The rules take into consideration the syntax of the functions (in  column 3 of the table) in order to convey: a sensitive sink (line 4), the sanitization (line 5), the validation (line 6), the extraction and modification (lines 7-10), and the concatenation (line 11).

\begin{lstlisting}[numbers=left, numbersep=5pt, caption=Grammar rules of ISL., label=f:stl_rules]
grammar isl {
slice-isl : statement+
statement : 
sensitive_sink
| sanitization
| validation
| mod_all 
| mod_add
| mod_sub 
| mod_rep 
| concat 
| cond statement+ cond? 
| assignment
sensitive_sink : ss (param | concat)
sanitization : sanit_f param
validation : (typechk_str | typechk_num | fillchk | contentchk) param
mod_all : (join_str | erase_str | replace_str | split_str) param 
mod_add : add_str param num_chars param
mod_sub : sub_str param num_chars start_where?
mod_rep : sub_str_replace param num_chars param start_where?
concat : (statement | param) (conc concat)?    
assignment : (statement | param) attrib_var
param : input | var     
attrib_var : var
num_chars : char5 | char6
}
\end{lstlisting}

As we will see in Section~\ref{s:hmm_ta}, tokens will correspond to the observations of the HMM. However, while a PHP assignment sets the value of the right-hand-side expression to the left-hand side, the tokens will be processed from left to right by the model; therefore, the assignment rule in ISL follows the HMM scheme.

\begin{example}
	PHP instruction \textcode{\$u = \$\_GET['user'];} is translated to \textcodeit{input var}. The assignment and parameter rules (lines 13, 22 and 23) derive the \textcodeit{input} token, while the attribution rule produces the \textcodeit{var} token (line 24).
\end{example}

\section{The Sequence Model}
\label{s:hmm_ta}

This section presents the \emph{sequence model} that supports vulnerability detection. It explains the graph that represents the model, identifying the states and the observations that can be emitted.

\subsection{Hidden Markov Model}
\label{s:hmm}

A Hidden Markov Model (HMM) is a statistical generative model that represents a process as a Markov chain with unobserved (hidden) states.
It is a dynamic Bayesian network with nodes that stand for random variables and edges that denote probabilistic dependencies between these variables \cite{Baum:66, Jurafsky:08, Smith:11}.
The variables are divided in two groups:  observed variables -- \emph{observations} -- and  hidden variables -- \emph{states}.
A state transitions to other states with some probability and  emits observations (see example in Fig.~\ref{f:seqs_corpus}). 

A HMM is specified by the following:
(1) a \emph{vocabulary}, a set of words, symbols or tokens that make up the sequence of observations;
(2) the \emph{states}, a group of states that classify the observations of a sequence;
(3) \emph{parameters}, a set of probabilities where (i) the initial probabilities indicate the probability of a sequence of observations begins at each start-state; (ii) the transition probabilities between states; and (iii)  the emission probabilities, which specify the probability of a state emitting a given observation.

In the context of NLP, sequence models are used to classify a series of observations, which correspond to the succession of words observed in a sentence. In particular, a HMM is used in PoS tagging tasks, allowing the discovery of a series of states that best explains a new sequence of observations. This is known as the \emph{decoding problem}, which can be solved by the Viterbi algorithm~\cite{Viterbi:67}. This algorithm resorts to dynamic programming to pick the best hidden state sequence. Although the Viterbi algorithm employs \emph{bigrams} to generate the \emph{i-th} state, it takes into account all previously generated  states, but this is not directly visible. In a nutshell, the algorithm iteratively obtains the probability distribution for the \emph{i-th} state based on the probabilities computed for the \emph{(i-1)-th} state, taking into consideration the parameters of the model.

The parameters of the HMM are \emph{learned} by processing a corpus that is created for training. Observations and state transitions are counted, and afterwards the counts are normalized in order to obtain probability distributions; a smoothing procedure may also be applied to deal with rare events in the training data (e.g., add-one smoothing).

\begin{table}[ht]\scriptsize
	\caption{\small HMM states and the observations they emit.}
	\label{t:states_emissions}
	\vspace{-3mm}	
	\centering					
	\begin{tabular}{|l|l|l|}					
		\hline					
		\textbf{State}	&	\textbf{Description}	&	\textbf{Emitted observations}	\\
		\hline\hline					
		
		Taint	&	Tainted	&	conc, input, var, var\_vv 	\\ 
		N-Taint	&	Not tainted	&	conc, cond, input, var, var\_vv, ss	\\ 
		San	&	Sanitization	&	input, sanit\_f, var, var\_vv	\\ 
		Val	&	Validation	&	contentchk, fillchk, input, typechk\_num,	\\
		&		&	typechk\_str, var, var\_vv	\\ 
		Chg\_str	&	Change string	&	add\_str, char5, char6, erase\_str,	input, \\
		&		&	  join\_str, replace\_str, split\_str, start\_where, 	\\
		&		&	sub\_str, sub\_str\_replace,   var, var\_vv	\\
		
		\hline	
	\end{tabular}
	\vspace{-4mm}
\end{table}

\subsection{Vocabulary and States}
\label{s:hmm_ta-voc}

As our HMM operates over the program instructions translated into ISL, the vocabulary is composed of the previously described ISL tokens. The states are selected to represent the fundamental operations that can be performed on the input data as it flows through a slice. Five states were defined as displayed Table \ref{t:states_emissions}.
The final state of an instruction in ISL is either vulnerable (\textcodeit{Taint}) or not-vulnerable (\textcodeit{N-Taint}). However, in order to attain an accurate detection, it is necessary to take into account the sanitization (\textcodeit{San}), validation (\textcodeit{Val}) and modification (\textcodeit{Chg\_str}) of the user inputs and the variables that may depend on them. Therefore, these three factors are represented as intermediate states in the model.
As strings are on the base of web surface vulnerabilities, these three states allow the model to determine the intermediate state when an application manipulates them.

\begin{figure}[b]
	\begin{center}
		\includegraphics[width=0.75\columnwidth]{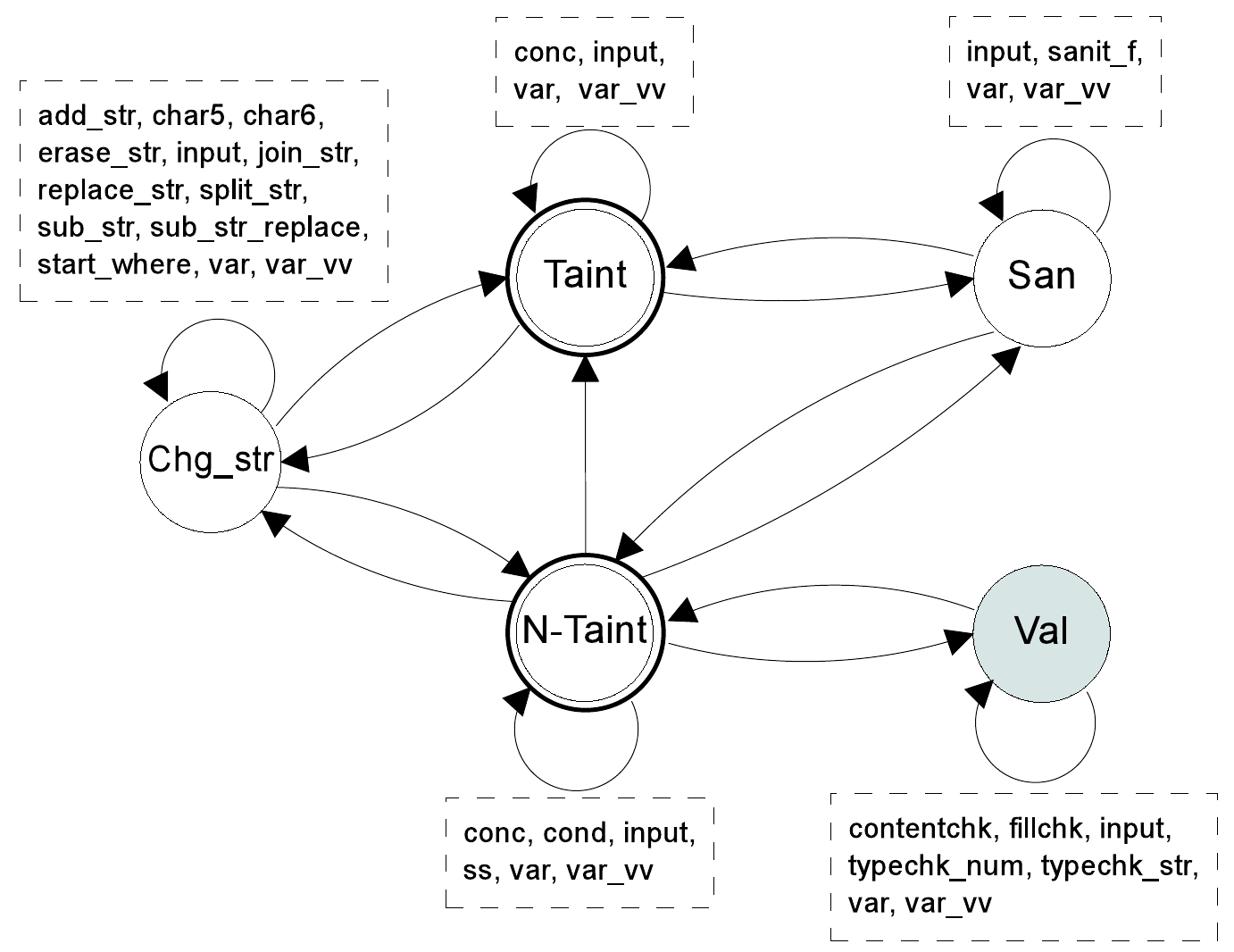}
	\end{center}
	\caption{\small Model graph of the proposed HMM.}
	\label{f:hmm_graph}
\end{figure}

\subsection{Graph of the Model}
\label{s:hmm_ta-graph}

Our HMM consists of the graph in Fig.~\ref{f:hmm_graph}, where the nodes constitute the states and the edges the transitions between them. The dashed squares next to the nodes hold the observations that can be emitted in each state.

An ISL instruction corresponds to a sequence of observations. The sequence can start in any state except \textcodeit{Val}. However, it can reach the \textcodeit{Val} state for example due to conditionals that check the input data. In the example of Fig.~\ref{t:xss} (b), in line 3, one notices a sequence that initiates with a \textcodeit{cond} observation that could be emitted by the \textcodeit{N-Taint}  initial state. Then, it would transit to the \textcodeit{Val} state due to the check that is carried out in the \textcodeit{if} conditional. When the processing of the sequence completes, the model is always either in the \textcodeit{Taint} or \textcodeit{N-Taint} states. Therefore, the final state determines the overall classification of the statement, i.e., if the instruction is vulnerable or not.

\begin{figure}[h] 
	\begin{flushleft} 
		\begin{center}
			\includegraphics[width=0.55\columnwidth]{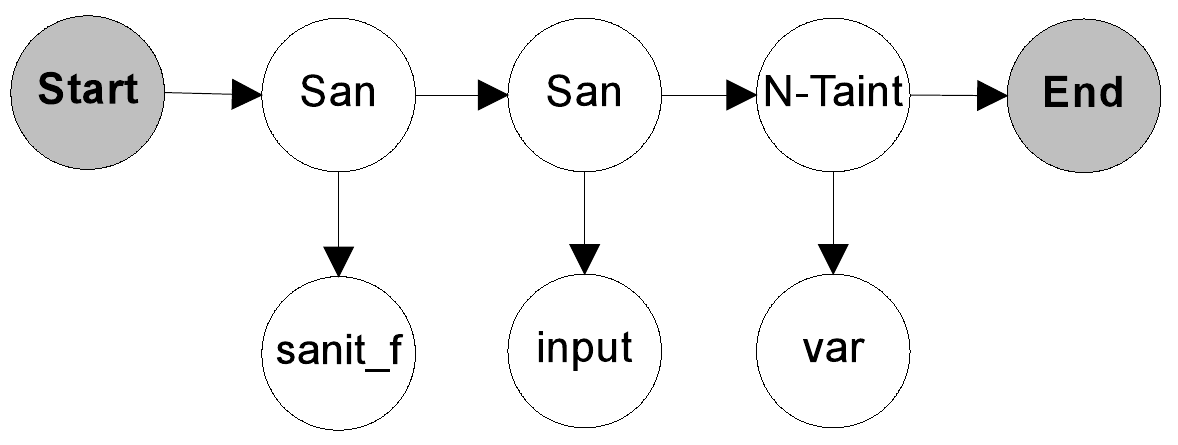}
		\end{center}
		\hspace{0.92cm}{\scriptsize(a) PHP instruction:} \textcode{\scriptsize \$p = mysqli\_real\_escape\_string(\$con, \$\_GET['user'])} \\ \vspace{-0.1cm}
		\hspace{1.3cm}{\scriptsize ISL instruction:} \textcode{\scriptsize sanit\_f input var} \\ \vspace{-0.1cm}
		\hspace{1.2cm} {\scriptsize Sequence:} \textcode{\scriptsize $\langle$sanit\_f,San$\rangle$ $\langle$input,San$\rangle$ $\langle$var,N-Taint$\rangle$}
		
	\end{flushleft}
	
	\caption{\small Graph instantiation for an example sequence.}
	\label{f:seqs_corpus}
\end{figure}

\begin{example}
	Fig.~\ref{f:seqs_corpus} shows an instantiation of the model for one sequence. The sanitization instruction is translated to the ISL sequence  \textcodeit{sanit\_f input var}. The sequence starts in the \textcodeit{San} state and emits the \textcodeit{sanit\_f} observation; next it remains in the same state and emits the \textcodeit{input} observation; then, it transits to \textcodeit{N-Taint} state, emitting the \textcodeit{var} observation (untainted variable).
\end{example}

\section{Learning and Vulnerability Detection}
\label{s:approach}

This section explains the main activities related with our approach. The learning phase encompasses a number of activities that culminate with the computation of the parameters of the HMM model. Following that, the vulnerabilities are found by processing the slices of the target application through model in the detection phase. Fig.~\ref{f:approach_arch_2} illustrates the fundamental steps.

\subsection{Slice Extraction and Translation Process}
\label{s:stl-translate}

The slice extractor analyses files with the source code, gathering the slices that start with an entry point and eventually reach some security-sensitive sink. The instructions between these points are those that implement the application logic based on the user input data. The slice extractor performs intra- and inter-procedural analysis, as it tracks the inputs and its dependencies along the program, walking through the invoked functions. The analysis is context-sensitive as it takes into account the results of function calls. 

A translation process occurs when the instructions are collected and consists in representing them as ISL tokens. However, ISL does not maintain much information about the variables portrayed by the \textcodeit{var} token. This knowledge is nevertheless crucial for a more accurate vulnerability detection as variables are related to the inputs in distinct manners and their contents can suffer all sorts of modifications. Therefore, to address this issue, we update a data structure called \emph{variable map} while the slice is translated. The map associates each occurrence of \textcodeit{var} in the \emph{slice-isl} with the name of the variable that appears in the source code. This lets us track how input data propagates to different variables when the slice code elements are processed.

There is an entry in the variable map per instruction. Each entry starts with a flag, 1 or 0, indicating if the statement is an assignment or not. The rest of the entry includes one value per token of the instruction, which is either the name of the variable (without the \textcode{\$}) or the \textcode{-} character (stands for a token that is not occupied by a variable).

\begin{example}
	Fig. \ref{t:classif_sqli_1}(a) displays a PHP code snippet that is vulnerable to SQLI and Fig. \ref{t:classif_sqli_1}(b) shows the translation into ISL and the variable map (ignore the right-hand side for now). 
	The first line is the assignment of an input to a variable, \textcodeit{\$u = \$\_POST['username'];}. As explained above, it becomes \textcodeit{input var} in ISL. The variable map entry \textcode{1 - u} is initialized to \textcode{1} to denote that the instruction is an assignment to the \textcode{var} in the second position.
	The next line is an assignment of a SQL query composed by concatenating constant substrings with a variable. It is represented in ISL by \textcodeit{var var} and in the variable map by \textcodeit{1 u q}. The last line corresponds to a sensitive sink (\textcodeit{ss}) and two variables.
\end{example}

\begin{example}
	Fig.~\ref{t:xss} has a slightly more complex code snippet. The slice extractor takes from the code two slices: lines \{1, 2, 3, 4\} and \{1, 3, 5, 6\}. The first prevents an attack with a form of input validation, but the second is vulnerable to XSS. The corresponding ISL and variable map are shown in the middle columns.
	The interesting cases are in lines 3 and 4, which are the \textcodeit{if} statement and its true branch. Both are prefixed with the  \textcodeit{cond} token and the former also ends with the same token. 
	This \textcodeit{cond} termination makes a distinction between the two types of instructions. In addition, the sequence model will understand that variables from the former may influence those that  appear in latter instructions.
\end{example}

\subsection{Process of Creating the Corpus}
\label{s:corpus}

The corpus plays an important role as it incorporates the knowledge that will be learned by the model, namely which instructions may lead to a flaw. In our case, the \emph{corpus} is a group of instructions (not slices) converted to ISL, where tokens are tagged with information related to taint propagation.  The model sees the tokens of an instruction in ISL as a sequence of observations. The tags correspond to the states of the model. Therefore, an alternative way to look at the corpus is as a group of sequences of observations annotated with states.

The corpus is built in four steps: 
(1) \emph{collection} of a group of instructions that are vulnerable and not-vulnerable, which are placed in a bag; 
(2) \emph{representation} of each instruction in the bag in ISL; 
(3) \emph{annotation} of the tokens of every instruction (e.g., as tainted or sanitized), i.e., associate a state to each observation of the sequence; and 
(4) \emph{removal} of duplicated entries in the bag. In the end, an instruction becomes a list of pairs of \textcode{$\langle$token,state$\rangle$}.

In the first step, it is necessary to get representative instructions of all classes of bugs that one wants to catch, various forms of validations, diverse forms of manipulating (changing) strings, and different combinations of code elements. To achieve this in practice, we can gather individual instructions or/and we can select a large number of slices captured from open source training applications. Therefore, both the collection and representation can be performed in an automatic manner (with the slice collector module), but the annotation of the tokens is done manually (as in all supervised machine learning approaches).

\begin{example}\label{ex:post}
	Instruction \textcode{\$var = \$\_POST['paramater']}  becomes \textcodeit{input var} in ISL, and is annotated as \textcode{$\langle$input,Taint$\rangle$ $\langle$var\_vv,Taint$\rangle$}. Both states are \textcodeit{Taint} (compromised) because the \textcodeit{input} can be the source of malicious data, and therefore is always \textcodeit{Taint}, and then the taint propagates to the variable. 
\end{example}

As mentioned in the previous section, the token \textcode{var\_vv} is not produced when slices are translated into ISL, but used in the corpus to represent variables with state \textcodeit{Taint} (tainted variables). In fact, during translation into ISL variables are not known to be tainted or not, so they are represented by the  \textcode{var} token. In the corpus, if the state of the variable is annotated as \textcodeit{Taint}, the variable is portrayed by \textcode{var\_vv}, forming the pair \textcode{$\langle$var\_vv,Taint$\rangle$}.

The state of the last observation of a sequence corresponds to a final state, and therefore it can only be \textcodeit{Taint} (vulnerable) or \textcodeit{N-Taint} (not-vulnerable). If this state is tainted then it means that a malicious input is able to propagate and potentially compromise the execution. Therefore, in this case, the instruction is perceived as vulnerable. Otherwise, the instruction is deemed correct (non-vulnerable).

\begin{example}\label{ex:htlmentities}
	Instruction \textcode{\$v = htlmentities (\$\_GET['user'])} is translated to \textcodeit{sanit\_f input var} and placed in the corpus as the succession of pairs \textcode{$\langle$sanit\_f,San$\rangle$ $\langle$input,San$\rangle$ $\langle$var,N-Taint$\rangle$}. The first two tokens are annotated with the \textcodeit{San} state because  function \textcode{htlmentities} sanitizes its parameter; the last token is labeled with the \textcodeit{N-Taint} state, meaning that the ultimate state of the sequence is not tainted. 
\end{example}

\begin{figure}[htb]
\begin{lstlisting}[breaklines=true, numbersep=5pt1]
$var = $_POST[`parameter']
$var = $_GET[`parameter']
$var = htmlentities($_POST[`parameter'])
$var = mysqli_real_escape_string($con, $_GET[`parameter'])
$var = htmlentities($var)
$var = "SELECT field FROM table WHERE field = $var"
$var = mysqli_query($con, $var)
$var = mysql_query($var)
echo $var
include($var)
$var = (isset($var)) ? $var : ''
if (isset($var) && $var > number)
if (is_string($var) && preg_match('pattern', $var))
if (isset($var) && preg_match('pattern', $var) && is_int($var))
\end{lstlisting}
 \vspace{-0.25cm}
	\captionof{lstlisting}{\small Creating the corpus: collection step.}
	\label{f_app:build_corpus_1}
\end{figure}

\begin{figure}
\begin{lstlisting}[breaklines=true, firstnumber=1, frame={top}, numbersep=5pt]
$var = $_POST[`parameter']
\end{lstlisting}\vspace{-4mm}
\begin{lstlisting}[breaklines=true, numbers=none, frame=none, numbersep=5pt]
input var_vv
\end{lstlisting}\vspace{-4mm}
\begin{lstlisting}[breaklines=true, firstnumber=2, frame=none, numbersep=5pt]
$var = $_GET[`parameter']
\end{lstlisting}\vspace{-4mm}
\begin{lstlisting}[breaklines=true, numbers=none, frame=none, numbersep=5pt]
input var_vv
\end{lstlisting}\vspace{-4mm}
\begin{lstlisting}[breaklines=true, firstnumber=3, frame=none, numbersep=5pt]
$var = htmlentities($_POST[`parameter'])
\end{lstlisting}\vspace{-4mm}
\begin{lstlisting}[breaklines=true, numbers=none, frame=none, numbersep=5pt]
sanit_f input var
\end{lstlisting}\vspace{-4mm}
\begin{lstlisting}[breaklines=true, firstnumber=4, frame=none, numbersep=5pt]
$var = mysqli_real_escape_string($con, $_GET[`parameter'])
\end{lstlisting}\vspace{-4mm}
\begin{lstlisting}[breaklines=true, numbers=none, frame=none, numbersep=5pt]
sanit_f input var
\end{lstlisting}\vspace{-4mm}
\begin{lstlisting}[breaklines=true, firstnumber=5, frame=none, numbersep=5pt]
$var = htmlentities($var)
\end{lstlisting}\vspace{-4mm}
\begin{lstlisting}[breaklines=true, numbers=none, frame=none, numbersep=5pt]
sanit_f var var
sanit_f var_vv var
\end{lstlisting}\vspace{-4mm}
\begin{lstlisting}[breaklines=true, firstnumber=6, frame=none, numbersep=5pt]
$var = "SELECT field FROM table WHERE field = $var"
\end{lstlisting}\vspace{-4mm}
\begin{lstlisting}[breaklines=true, numbers=none, frame=none, numbersep=5pt]
var var
var_vv var_vv
\end{lstlisting}\vspace{-4mm}
\begin{lstlisting}[breaklines=true, firstnumber=7, frame=none, numbersep=5pt]
$var = mysqli_query($con, $var)
\end{lstlisting}\vspace{-4mm}
\begin{lstlisting}[breaklines=true, numbers=none, frame=none, numbersep=5pt]
ss var var
ss var_vv var_vv
\end{lstlisting}\vspace{-4mm}
\begin{lstlisting}[breaklines=true, firstnumber=8, frame=none, numbersep=5pt]
$var = mysql_query($var)
\end{lstlisting}\vspace{-4mm}
\begin{lstlisting}[breaklines=true, numbers=none, frame=none, numbersep=5pt]
ss var var
ss var_vv var_vv
\end{lstlisting}\vspace{-4mm}
\begin{lstlisting}[breaklines=true, firstnumber=9, frame=none, numbersep=5pt]
echo $var
\end{lstlisting}\vspace{-4mm}
\begin{lstlisting}[breaklines=true, numbers=none, frame=none, numbersep=5pt]
ss var_vv
ss var
\end{lstlisting}\vspace{-4mm}
\begin{lstlisting}[breaklines=true, firstnumber=10, frame=none, numbersep=5pt]
include($var)
\end{lstlisting}\vspace{-4mm}
\begin{lstlisting}[breaklines=true, numbers=none, frame=none, numbersep=5pt]
ss var_vv
ss var
\end{lstlisting}\vspace{-4mm}
\begin{lstlisting}[breaklines=true, firstnumber=11, frame=none, numbersep=5pt]
$var = (isset($var)) ? $var : ''
\end{lstlisting}\vspace{-4mm}
\begin{lstlisting}[breaklines=true, numbers=none, frame=none, numbersep=5pt]
var var
var_vv var_vv
\end{lstlisting}\vspace{-4mm}
\begin{lstlisting}[breaklines=true, firstnumber=12, frame=none, numbersep=5pt]
if (isset($var) && $var > number)
\end{lstlisting}\vspace{-4mm}
\begin{lstlisting}[breaklines=true, numbers=none, frame=none, numbersep=5pt]
cond fillchk var_vv cond
cond fillchk var cond
\end{lstlisting}\vspace{-4mm}
\begin{lstlisting}[breaklines=true, firstnumber=13, frame=none, numbersep=5pt]
if (is_string($var) && preg_match('pattern', $var))
\end{lstlisting}\vspace{-4mm}
\begin{lstlisting}[breaklines=true, numbers=none, frame=none, numbersep=5pt]
cond typechk_str var_vv contentchk var_vv cond
cond typechk_str var_vv contentchk var cond
cond typechk_str var contentchk var_vv cond
cond typechk_str var contentchk var cond
\end{lstlisting}\vspace{-4mm}
\begin{lstlisting}[breaklines=true, firstnumber=14, frame=none, numbersep=5pt]
if (isset($var) && preg_match('pattern', $var) && is_int($var))
\end{lstlisting}\vspace{-4mm}
\begin{lstlisting}[breaklines=true, numbers=none, frame=none, numbersep=5pt]
cond typechk_str var_vv contentchk var_vv typechk_int var_vv cond
cond typechk_str var_vv contentchk var_vv typechk_int var cond
cond typechk_str var_vv contentchk var typechk_int var_vv cond
cond typechk_str var_vv contentchk var typechk_int var cond   
cond typechk_str var contentchk var_vv typechk_int var_vv cond
cond typechk_str var contentchk var_vv typechk_int var cond
cond typechk_str var contentchk var typechk_int var_vv cond
cond typechk_str var contentchk var typechk_int var cond
\end{lstlisting}    
	\vspace{-0.1cm}
	\hrule
    \vspace{-0.25cm}
	\captionof{lstlisting}{\small Creating the corpus: representation step.}
	\label{f_app:build_corpus_2}
\end{figure}

\begin{figure}[htb]
\begin{lstlisting}[breaklines=true, numbersep=5pt] 
<input,Taint> <var_vv,Taint>
<sanit_f,San> <input,San> <var,N-Taint>
<sanit_f,San> <var,San> <var,N-Taint>
<sanit_f,San> <var_vv,San> <var,N-Taint>
<var,N-Taint> <var,N-Taint>
<var_vv,Taint> <var_vv,Taint>
<ss,N-Taint> <var,N-Taint> <var,N-Taint>
<ss,N-Taint> <var_vv,Taint> <var_vv,Taint>
<ss,N-Taint> <var_vv,Taint>
<ss,N-Taint> <var,N-Taint>
<cond,N-Taint> <fillchk,Val> <var_vv,Val> <cond,N-Taint>
<cond,N-Taint> <fillchk,Val> <var,Val> <cond,N-Taint>
<cond,N-Taint> <typechk_str,Val> <var_vv,Val> <contentchk,Val> <var_vv,Val> <cond,N-Taint>
<cond,N-Taint> <typechk_str,Val> <var_vv,Val> <contentchk,Val> <var,Val> <cond,N-Taint>
<cond,N-Taint> <typechk_str,Val> <var,Val> <contentchk,Val> <var_vv,Val> <cond,N-Taint>
<cond,N-Taint> <typechk_str,Val> <var,Val> <contentchk,Val> <var,Val> <cond,N-Taint>
<cond,N-Taint> <typechk_str,Val> <var_vv,Val> <contentchk,Val> <var_vv,Val> <typechk_int,Val> <var_vv,Val> <cond,N-Taint>
<cond,N-Taint> <typechk_str,Val> <var_vv,Val> <contentchk,Val> <var_vv,Val> <typechk_int,Val> <var,Val> <cond,N-Taint>
<cond,N-Taint> <typechk_str,Val> <var_vv,Val> <contentchk,Val> <var,Val> <typechk_int,Val> <var_vv,Val> <cond,N-Taint>
<cond,N-Taint> <typechk_str,Val> <var_vv,Val> <contentchk,Val> <var,Val> <typechk_int,Val> <var,Val> <cond,N-Taint>
<cond,N-Taint> <typechk_str,Val> <var,Val> <contentchk,Val> <var_vv,Val> <typechk_int,Val> <var_vv,Val> <cond,N-Taint>
<cond,N-Taint> <typechk_str,Val> <var,Val> <contentchk,Val> <var_vv,Val> <typechk_int,Val> <var,Val> <cond,N-Taint>
<cond,N-Taint> <typechk_str,Val> <var,Val> <contentchk,Val> <var,Val> <typechk_int,Val> <var_vv,Val> <cond,N-Taint>
<cond,N-Taint> <typechk_str,Val> <var,Val> <contentchk,Val> <var,Val> <typechk_int,Val> <var,Val> <cond,N-Taint>
\end{lstlisting}
	\vspace{-0.4cm}
	\captionof{lstlisting}{\small Creating the corpus: annotation and removal steps.}
	\label{f_app:build_corpus_3}
\end{figure}

\begin{example}
	Listing \ref{f_app:build_corpus_1} displays fourteen PHP instructions collected from vulnerable and non-vulnerable slices. The representation of the instructions into ISL is illustrated in Listing \ref{f_app:build_corpus_2}. 
	It is possible to observe that some instructions may have more than one representation, depending if the extracted slice is vulnerable or not. For example, the instruction fifth position in Listing \ref{f_app:build_corpus_2} appears as two series (the two lines immediately below of it) corresponding to the sanitization of an untainted and a tainted variable, respectively. 
	In the listing, it is also visible the difference between the \textcode{var} and \textcode{var\_vv} tokens. 
	Listing \ref{f_app:build_corpus_3} has the final corpus that is produced after applying the last two steps. Each sequence of observations is annotated with the state as explained above. The duplicated sequences are eliminated as several PHP instructions can result in the same sequence. For example, PHP instructions in lines 1 and 2 (Listing \ref{f_app:build_corpus_1}) become the same sequence (line 1 of Listing \ref{f_app:build_corpus_3}). 
\end{example}

\subsection{Configuring the HMM}
\label{s:hmm_ta-param}

The sequence model was mostly defined in Section~\ref{s:hmm_ta}. The only missing piece of information are the \emph{parameters}, i.e., the various probabilities to arrive to the start-states, to do the state transitions, and to perform the emissions of the observations. The probabilities are computed from the corpus by counting the number of occurrences of observations and/or states. The result is 3 matrices of probabilities with dimensions of $(1\times s)$, $(s\times s)$ and $(t\times s)$, where $s$ and $t$ are the number of states and tokens of the model.  
The matrices are calculated as follows:

\noindent
\emph{Start-state probabilities:} count how many sequences begin in each state. Then, get the probability for each state by dividing these counts by the number of sequences of the corpus. This produces a matrix with the dimension $(1\times 5)$.
\begin{example}
	To obtain the start-state probability of the \textcodeit{San} state, we  count how many sequences begin with the \textcodeit{San} state and divide by the size of the \emph{corpus}.
\end{example}

\noindent
\emph{Transition probabilities:} count how many times in the corpus a certain state $i$ transits to a state $k$ (including itself). The transition probability is obtained by dividing this count by the number of pairs of states that appear in the corpus that begin with the $i$ state.
The resulting matrix has a dimension of $(5\times 5)$, keeping the various probabilities for all possible transitions between the five states.

\begin{example}
	The transition probability for the \textcodeit{N-Taint} state to the \textcodeit{Taint} state is the number of occurrences of this transition in the corpus divided by the number of pairs of states that begin in the \textcodeit{N-Taint} state.
	
\end{example}

\noindent
\emph{Emission probabilities:} count how many times a certain observation is emitted by a particular state, i.e., count how many times a certain pair \textcode{$\langle$token,state$\rangle$} appears in the corpus. Then, calculate the emission probability by dividing this count by the total number of pairs \textcode{$\langle$token,state$\rangle$} that occur for that specific state.
The resulting matrix -- called \emph{global emission probabilities matrix} -- has a dimension of $(22\times 5)$ in order to have a probability for the 22 tokens that could emitted by each of the 5 states.

\begin{example}
	To obtain the probability that the \textcodeit{Taint} state  emits the \textcodeit{var\_vv} token (\textcode{$\langle$var\_vv,Taint$\rangle$}), first get the number of occurrences of this pair in the corpus, and next divided it by the total number of pairs of the \textcodeit{Taint} state.
\end{example} 

Zero-probabilities should be avoided because the Viterbi algorithm uses multiplication to calculate the probability of moving to the next state, and therefore one needs to ensure that this multiplication is never zero.
The \emph{add-one smoothing} technique \cite{Jurafsky:08} can address this issue and help to compute the values of the parameters. This technique simply adds a unit to all counts, making zero-counts equal to one and the associated probability different from zero.

\subsection{Detecting Vulnerabilities}
\label{s:hmm_ta-classify}

Given the source code of an application, the collector gathers the slices that should be examined, and then every slice is inspected separately. To commence, the instructions of the slice are translated to ISL. This means that the slice becomes a list of sequences of observations, each one corresponding to a PHP instruction. The discovery of flaws is accomplished by processing the sequences in the order of appearance, starting with the first and concluding with the last.

The HMM model is applied to each sequence of observations to find out the associated states. 
We resort to an extension of the Viterbi algorithm to perform this task. The algorithm employs dynamic programming to compute the most likely succession of states that explain a sequence of observations. As the algorithm finishes with a sequence, a final state comes out, either as \textcodeit{Taint} or \textcodeit{N-Taint}. This information is then propagated to the next sequence. The process is repeated for all sequences, and the final state of the last sequence defines the outcome for the slice --- either as vulnerable (if it is tainted) or non-vulnerable (if it is untainted).

For the classification to be carried out effectively, it is necessary to spread faithfully the taintedness among the sequences under analysis, which means keeping information about the variables that are tainted. For this purpose,  we use three artifacts that are updated as the execution evolves: 

\begin{itemize}
	\item \emph{Tainted List} (TL): as sequences are processed, it keeps the identifiers of the variables that are perceived as tainted;
	
	\item \emph{Conditional Tainted List} (CTL): contains the inputs (token \textcodeit{input}) and tainted variables (belong to TL) that have been validated (e.g., by tokens \textcodeit{typechk\_num} and \textcodeit{contentchk});
	
	\item \emph{Sanitized List} (SL): has essentially a similar aim as CTL, except that it maintains the variables that are sanitized or modified (e.g., with functions that manipulate strings).
\end{itemize}

\begin{example}
	Fig.~\ref{t:xss} has the PHP code for the two slices composed of lines \{1, 2, 3, 4\} and \{1, 3, 5, 6\} respectively. 
	After processing the first slice, TL = \{u, a\} and CTL = \{u, a\} as variable $u$ is the parameter of the \textcodeit{contentchk} token and variable $a$ is the parameter of the \textcodeit{typechk\_int} token. The final state is \textcodeit{N-Taint} because variable $u$ is included in CTL. In the other slice, TL = \{u, a\} and CTL = \{ \} since there is no validation and the final state is \textcodeit{Taint}.
\end{example}

In our implementation, the Viterbi algorithm was extended to explore the information kept in the variable map and in these artifacts (further details in Section \ref{ss:mod_viterbi}). Handling a sequence of observations becomes a three step procedure: (1) a preprocessing step is carried out -- \textcodeit{beforeVit}; (2) then, the decoding step of the Viterbi algorithm is applied -- \textcodeit{decodeVit}; (3) and lastly, a post-processing step is executed -- \textcodeit{afterVit}.
They work as follows:

\begin{description} 
	\item [\textcodeit{beforeVit}:] the variable map is visited to get the name of the variable associated to each \textcodeit{var} observation. The TL and SL are checked to determine if they hold that name. In case the sequence starts with the token \textcodeit{cond}, the list CTL is also accessed. If a variable only belongs to TL, then the \textcodeit{var} observation is modified to \textcodeit{var\_vv}, thus capturing the effect of the variable being tainted. Finally, an emission probability sub-matrix for the observations of the sequence is also retrieved from the global emission probabilities matrix;
	
	\item [\textcodeit{decodeVit}:] for each observation, the Viterbi algorithm calculates the probability of each state to emit it, considering the probabilities of emission, of transition, and of the states already discovered. The multiplication of these three probabilities results in a probability called \emph{score of state}. The state that is assigned to an observation is the one that has the highest score. The process is repeated for all observations and the state of the last observation is the one that classifies the sequence as \textcodeit{Taint} or \textcodeit{N-Taint}. 
	
	In more detail, the three probabilities are obtained as follows: emission come from the sub-matrix of emission probabilities, regarding the observations that will be processed; transition are from the matrix of transition probabilities; previous state is determined by picking up the highest score computed for the previous observation. This last probability brings to the calculation the order in which the observations appear in the sequence and the knowledge already discovered about the previous observations. However, since this knowledge does not exist for the first observation of the sequence, in this case the start-state probabilities are used;
	
	\item [\textcodeit{afterVit}:] if the sequence \emph{is an assignment} (i.e., the last observation of the sequence is a \textcodeit{var} token and the entry in the variable map starts with 1), then the corresponding variable name is obtained from variable map. Next, the TL is updated: (i) inserting the variable name if the final state is \textcodeit{Taint}; or (ii) removing it if the state is \textcodeit{N-Taint} and the variable is in TL; in the presence of a sanitization sequence, the variable name is also added to SL. In case the sequence \emph{is an \textcodeit{if} condition} (i.e., the first and last observations are a \textcodeit{cond} token), then the variable map is searched for each \textcodeit{var} and \textcodeit{var\_vv} observation. Next, the TL is searched to discover if it includes the name, and in that situation, the CTL is updated by inserting that name. 
\end{description}

The end result of these actions is that one gets the ability to keep the relevant knowledge about the propagation of inputs through the slice, and thus determine how they can influence the sensitive sinks.

\begin{example}
	Fig.~\ref{t:classif_sqli_1}(a) and (b) shows an example of the detection of a bug. It comprises from left to right: the PHP code, the representation in ISL, the variable map, and the TL after observations are classified.
	In line 1, the Viterbi algorithm is applied and as result the \textcodeit{var} observation is tainted because by default an \textcodeit{input} observation is so; the model classifies it correctly and variable \textcodeit{u} is inserted in TL. 
	In line 2, the first \textcodeit{var} observation is updated to \textcodeit{var\_vv} because it corresponds to variable \textcodeit{u} that belongs to TL, and then the Viterbi algorithm is applied; the \textcodeit{var\_vv var} sequence is classified by the model and the final state is \textcodeit{Taint}; therefore, variable \textcodeit{q} is inserted in TL. The process is repeated for the next line, allowing the discovery of the flaw. 
	Fig.~\ref{t:classif_sqli_1}(c) presents the decoding of the slice while the processing progresses. Here, it is possible to see the places where \textcodeit{var} is replaced by \textcodeit{var\_vv}, with the relevant variable name as suffix. In addition, the states of each observation are also added.  By following the generated states, one can understand the effects of the code execution (without actually running it), which variables are tainted, and why the code is vulnerable. The state of the last observation indicates the final classification --- a vulnerability.
\end{example}

\section{Implementation and Initial Assessment}
\label{s:impl_eval_dekant}

Our approach is implemented in the DEKANT tool. A corpus was also created to train the model. This corpus can be extended in the future with additional annotated sequences, allowing the tool to evolve its knowledge and detection capabilities.

\subsection{Implementation of DEKANT}
\label{ss:impl_dekant}
\label{subsub:vul_det}

DEKANT is programmed in Java and its architecture is divided in four major modules, which are explained below in more detail:

\noindent
\emph{Knowledge extractor:}  operates separately from the other modules and is executed when the corpus is built or later modified. It runs in three steps:
(i) the sequences composed of series of annotated tokens are loaded from a plain text file. Each sequence is separated in pairs \textcode{$\langle$token,state$\rangle$} and the elements of each pair are inserted in the matrices called \emph{observations} and \emph{states}. Since sequences normally have different numbers of pairs, it becomes necessary to \emph{normalize the length of all sequences} in the corpus. 
This is accomplished by first determining the length of the largest sequence, and then by padding shorter sequences with the \textcodeit{miss} token together with the state of the last observation (i.e., with pairs \textcode{$\langle$miss,Taint$\rangle$} or \textcode{$\langle$miss,N-Taint$\rangle$}) to ensure that all sequences have the same length; 
(ii) then, the various probabilities of the model are computed as explained in the previous section;
(iii) lastly, all relevant information about the model is saved in a plain text file to be loaded by the vulnerability detector module. 

\noindent 
\emph{Slice collector:} uses a lexer and a parser to process PHP code (based on ANTLR\footnote{https://www.antlr.org/}). It searches the application files for places where inputs arrive from the user and then tracks the data flows until either a security-critical instruction is reached or the program exits. 
Slices that have both an entry point and a sensitive sink are passed to the translator (and the others are discarded). 
The information about which entry points and sensitive sinks should be considered is provided in a configuration file. 

\noindent
\emph{Slice translator:} The module reads configuration files describing the classes of tokens, e.g., containing the PHP functions that are represented by tokens. Some of them are transversal to any class of vulnerability, whereas others are specific to a particular bug. For example, the \textcode{input} file contains \textcode{\$\_GET} and \textcode{\$\_POST} global arrays and the \textcode{ss\_xss} file has the security-sensitive functions associated with XSS (e.g., \textcode{echo}). The module first parses the slice and next verifies which tokens should be assigned to each PHP instruction, following the ISL grammar rules. Simultaneously, it also generates the variable map.

\noindent
\emph{Vulnerability detector:} works in three steps to find the bugs.
(i) the probabilities are loaded from a file and the model is setup internally; (ii) the slice translated into the intermediate language is processed using the modified Viterbi algorithm. Sometimes, it occurs that a sequence has more observations than the largest sequence that was seen in the corpus. When this happens, it is necessary to divide the sequence in sub-sequences with at most the maximum corpus sequence length. Then, each one is classified separately, but the algorithm is careful to ensure that the initial probability of the following sub-sequence is equal to the probability resulting from the previous sub-sequence; (iii) lastly,  the various probabilities are estimated for a sequence of observations to be explained by particular sequences of states, and the most probable is chosen. An alert message is issued if a vulnerability is found. 

\subsubsection{Extensions to the Viterbi algorithm}
\label{ss:mod_viterbi}

We extended the Viterbi algorithm with the two procedures of Section \ref{s:hmm_ta-classify} (\textcodeit{beforeVit} and \textcodeit{afterVit}) to track the propagation of inputs while processing a slice and to explore the data structures that keep relevant knowledge about variables (e.g., the three artifacts TL, CTL and SL).

	Listing \ref{f:viterbi_modified_1} presents the \textcodeit{beforeVit} preprocessing procedure that is run before the Viterbi algorithm. 
	\textcode{beforeVit} does a few tests to manipulate some flags and change the data structures. For each observation (\textcode{obs}) in the sequence (\textcode{inst\_slice\_isl}) there are checks to find out: (i) the presence of sanitization  (\textcode{sanit\_f}) or a \textcode{cond} tokens. For the latter case, it is verified the \textcode{obs} position in the sequence to discover if the instruction is an \textcode{if} statement, an instruction inside of a conditional statement, or an \textcode{else} statement (lines 19 to 30);
	(ii) an \textcode{if} statement is searched for validation functions and if their parameters are a variable or an input (i.e., \textcode{var} or \textcode{input}). In such case, \textcode{var} or \textcode{input} are inserted in CTL (lines 32 to 48);
	(iii) an instruction inside an \textcode{if} statement is checked if the \textcode{var} and \textcode{input} tokens  belong to CTL and/or SL. VM (variable map) is accessed to get the name of variable associated to \textcode{var} token. If the token \textcode{input} belongs to the SL or CTL lists, it is replaced by the \textcode{var} token because it has to loose its taintedness (we recall that by default this token is tainted and the \textcode{var} token is untainted, so this replacement is required) (lines 50 to 61);
	(iv) in  presence of another instruction and if the observation is a \textcode{var} token, i.e., the \textcode{inst\_slice\_isl} is out of the validation scope, the name of variable is taken from VM and checked if it belongs to TL but not in SL. In such a case, the variable is tainted, and the observation is replaced by the \textcode{var\_vv} token (lines 63 to 71).
	For all four verifications, the emission probability of the observation in analysis is retrieved from the global emission probabilities matrix (\textcode{GEP}), then it is inserted in the emission probabilities matrix (\textcode{EP}) of the \textcode{inst\_slice\_isl} (line 72).
	
	Afterwards, the traditional Viterbi algorithm is executed (\textcodeit{decodeVit} step as explained in Section \ref{s:hmm_ta-classify}) and then the post-processing \textcode{afterVit} procedure runs.

\begin{lstlisting}[numbers=left, numbersep=5pt, breaklines=true, language=pascal, caption={\small  {\emph{beforeVit}} extension to the Viterbi algorithm.}, captionpos=b, label=f:viterbi_modified_1] 
/* >>> Data structures and variables <<<
** VM - variable map
** TL - tainted list
** CTL - conditional tainted list
** SL - sanitized list
** obs_index - index of obs in the instruction_slice_isl
** var_name - variable name of the obs from inst_slice_isl
** condition - variable for controlling if stattements
** val - variable for controlling validation functions 
** san - variable for controlling sanitization functions
** EP - emission probability matrix of instruction_slice_isl
** GEP - global emission probabilities matrix
** obs_ep - emission probability of the obs in analysis
*/

val = 0
san = 0
for each obs in inst_slice_isl do
   if obs = sanit_f then san = 1 end_if

   if obs = cond then
      if obs_index = 1 then
         if size(inst_slice_isl) = 1 then condition = 0 else condition = 1 end_if
      else
         condition = 2
      end_if
      get obs_ep from GEP
   end_if

   if condition = 1 and obs_index <> 1 then
      if obs in [typechk_num, contentchk] then
         val = 1
      end_if

      if obs = var and val = 1 then
         get var_name of obs from VM
         insert var_name in CTL
         val = 0
      end_if

      if obs = input and val = 1 then
         insert input in CTL
         val = 0
      end_if
      get obs_ep from GEP
   end_if  

   if condition = 2 then
      if obs = var then
         get var_name of obs from VM
         if var_name in [CTL, SL] then 
            get obs_ep from GEP
         end_if
      end_if
      if obs = input and input in [CTL, SL] then
         obs = var
         get obs_ep from GEP
      end_if
   end_if

   if condition = 0 then
      if obs = var then
         get var_name of obs from VM
         if var_name in TL and not in SL then
            obs = var_vv
         end_if
      end_if
      get obs_ep from GEP
   end_if
   insert obs_ep in EP
end_do
\end{lstlisting}


	Listing \ref{f:viterbi_modified_2} shows \textcode{afterVit}. It takes as inputs the final state of the \textcode{inst\_slice\_isl} (\textcode{state}) and the assignment value (\textcode{value}) of the instruction stored in VM (lines 11-12), and then makes the following checks: 
	(i) if the instruction is an assignment, then the last observation of the sequence is a variable (\textcode{var} or \textcode{var\_vv} token), so the name of the variable (\textcode{var\_name}) is taken from VM (lines 14-15).
	(ii) if the instruction is classified as \textcodeit{Taint}, then the assignment variable is tainted, so the \textcode{var\_name} is put in TL. If this \textcode{var\_name} already belongs to SL, it is removed from this list (lines 16-20).
	(iii) if the instruction is classified as \textcodeit{N-Taint}, then the assignment variable is untainted, and therefore it can be removed from TL. Additionally, it is verified if the instruction is a result of a sanitization operation, and in such case the name is inserted in SL (lines 22-26).

\begin{lstlisting}[breaklines=true, numbersep=5pt, language=Pascal, caption={\small {\emph{afterVit}} extension to the Viterbi algorithm.}, captionpos=b, label=f:viterbi_modified_2] 
/* >>> Data structures and variables <<<
** VM - variable map
** TL - tainted list
** SL - sanitized list
** state - state of the last obs from inst_slice_isl
** value - assignament value of inst_slice_isl on VM 
** var_name - variable name of the obs from inst_slice_isl
** san - variable for controlling sanitization functions
*/

get state of inst_slice_isl
get value from VM

if value = 1 then
   get var_name of the last_obs from VM
   if state = taint then
      insert var_name in TL
      if var_name in SL then
         remove var_name from SL
      end_if
   else
      if san = 1 then
         insert var_name in SL
         san = 0
         if var_name in TL then
            remove var_name from TL
         end_if      
      end_if
   end_if
end_if
\end{lstlisting}

\subsection{Corpus Construction and Assessment}
\label{ss:eval_model}

The model needs to classify correctly the sequences of observations or, in our case, needs to detect vulnerabilities without mistakes. 
Since the model is configured with the corpus, its quality depends strongly on  incorporating valid and enough information in the corpus. Therefore, 
to build the corpus, we resorted to a method inspired in Jurafsky and Martin \cite{Jurafsky:08}. The method operates iteratively in three phases to gradually assess and improve the resulting model.  The \emph{evaluation phase}  verifies if the model outputs correctly a sequence of observations $O$ for a given sequence of states $S$.  The \emph{decoding phase}  determines if the model outputs a $S$ that explains correctly a given $O$. This phase corresponds to the objective of our approach. The last phase, \emph{re-learning}, verifies if the model needs adjustments to its parameters in order to maximize the results of the previous phases. It consists of enhancing the model by adding more sequences to the corpus and running another cycle of the method.

\begin{figure}[htb]
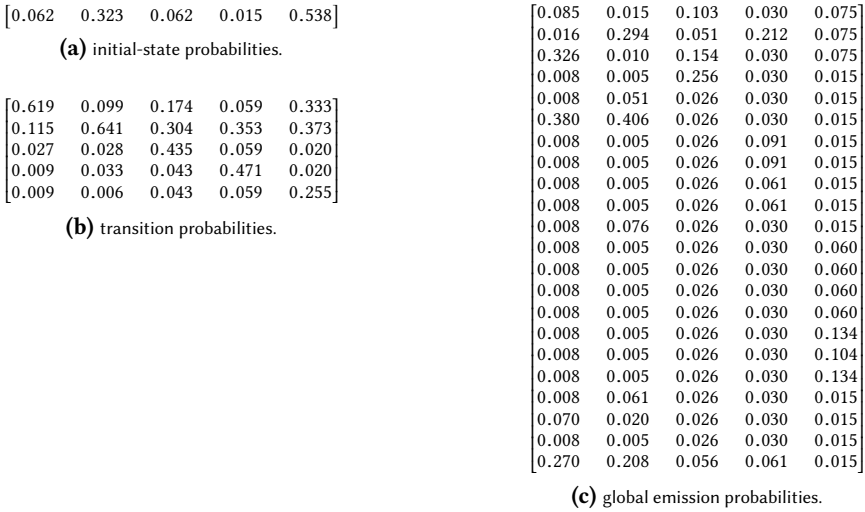
\scriptsize
		\centering 
		\begin{subfigure}[t]{0.5\columnwidth}  
			\begin{subfigure}[t]{1\columnwidth}
				\[
				\begin{bmatrix}
				0.062 & 0.323 & 0.062 & 0.015 & 0.538
				\end{bmatrix}
				\]
				\vspace{-0.4cm}
				\caption{\scriptsize initial-state probabilities.}
			\end{subfigure}%
			
			\begin{subfigure}[t]{1\columnwidth}
				\vspace{0.4cm}
				\[
				\begin{bmatrix}
				0.619 & 0.099 & 0.174 & 0.059 & 0.333 \\
				0.115 & 0.641 & 0.304 & 0.353 & 0.373 \\
				0.027 & 0.028 & 0.435 & 0.059 & 0.020 \\
				0.009 & 0.033 & 0.043 & 0.471 & 0.020 \\
				0.009 & 0.006 & 0.043 & 0.059 & 0.255
				\end{bmatrix}
				\]
				\vspace{-0.2cm}
				\caption{\scriptsize transition probabilities.}
			\end{subfigure}%
		\end{subfigure}%
		~
		\begin{subfigure}[t]{0.5\columnwidth}
			\[
			\begin{bmatrix}
			0.085 & 0.015 & 0.103 & 0.030 & 0.075 \\
			0.016 & 0.294 & 0.051 & 0.212 & 0.075 \\
			0.326 & 0.010 & 0.154 & 0.030 & 0.075 \\
			0.008 & 0.005 & 0.256 & 0.030 & 0.015 \\
			0.008 & 0.051 & 0.026 & 0.030 & 0.015 \\
			0.380 & 0.406 & 0.026 & 0.030 & 0.015 \\
			0.008 & 0.005 & 0.026 & 0.091 & 0.015 \\
			0.008 & 0.005 & 0.026 & 0.091 & 0.015 \\
			0.008 & 0.005 & 0.026 & 0.061 & 0.015 \\
			0.008 & 0.005 & 0.026 & 0.061 & 0.015 \\
			0.008 & 0.076 & 0.026 & 0.030 & 0.015 \\
			0.008 & 0.005 & 0.026 & 0.030 & 0.060 \\
			0.008 & 0.005 & 0.026 & 0.030 & 0.060 \\
			0.008 & 0.005 & 0.026 & 0.030 & 0.060 \\
			0.008 & 0.005 & 0.026 & 0.030 & 0.060 \\
			0.008 & 0.005 & 0.026 & 0.030 & 0.134 \\
			0.008 & 0.005 & 0.026 & 0.030 & 0.104 \\
			0.008 & 0.005 & 0.026 & 0.030 & 0.134 \\
			0.008 & 0.061 & 0.026 & 0.030 & 0.015 \\
			0.070 & 0.020 & 0.026 & 0.030 & 0.015 \\
			0.008 & 0.005 & 0.026 & 0.030 & 0.015 \\
			0.270 & 0.208 & 0.056 & 0.061 & 0.015
			\end{bmatrix}
			\]
			\vspace{-0.2cm}
			\caption{\scriptsize global emission probabilities.}
		\end{subfigure}
		\vspace{-0.3cm}
		\caption{\small Parameters of our HMM model extracted from the corpus. Columns correspond to the 5 states (in the same order of column 1 of Table \ref{t:states_emissions}). The lines of matrix (c) are the tokens (in the same order of column 1 of Table \ref{t:ttokens}).}
		\label{f:prob}
	\end{figure}

After applying the method, the resulting corpus had 510 slices, where 414 are vulnerable and 96 are non-vulnerable. These slices were extracted from various open source PHP applications\footnote{\scriptsize bayar, bayaran, ButterFly, CurrentCost, DVWA 1.0.7, emoncms, glfusion-1.3.0, hotelmis, Measureit 1.14, Mfm-0.13, mongodb-master, Multilidae 2.3.5, openkb.0.0.2, Participants-database-1.5.4.8, phpbttrkplus-2.2, SAMATE, superlinks, vicnum15, ZiPEC 0.32, Wordpress 3.9.1.} and had flaws from the twelve classes presented in Section \ref{prob:vvs}. The probability matrices that were computed based on this corpus are shown in
Fig.~\ref{f:prob}.

To perform a preliminary assessment the model, we applied a \emph{10-fold cross validation} \cite{Demsar:06}. This form of validation involves dividing the training data (the corpus of 510 slices) in ten folds. Then, the model is trained with a sub-corpus of nine of the folds and tested with the tenth fold. This process is repeated ten times to evaluate every fold with a model trained with the rest. The metrics that are used in the evaluation are: \emph{Accuracy (acc)} measures the ratio of well-classified slices  (as vulnerable and non-vulnerable) over the total number of slices (\emph{N}), whereas \emph{precision (pr)} assesses the fraction of classified bugs that are really vulnerabilities.
The objective is high accuracy and precision or, similarly, to minimize the \emph{false positive rate (fpr)} which is the rate of generating false alarms for slices that are correct, and to minimize the \emph{false negative rate (fnr)} which is the rate of missing certain vulnerable slices. 
Given that \emph{tp} and \emph{tn} are the well-classified instances as vulnerable and non-vulnerable, while \emph{fp} is the false alarms and \emph{fn} is the missing alarms, the metrics are computed with: $acc = (tp+tn)/N$; $pr = tp/(tp + fp)$; $fpr = fp/(fp + tn)$; and $fnr = fn/(fn + tp)$.

Table~\ref{t:corpus_cross_val} presents a confusion matrix for the alerts produced by DEKANT in the first two  phases of the method. For example, the first row says that DEKANT issued 419 alerts in the evaluation phase but that 14 of them were mistakes (columns 2 and 3).  In the evaluation phase, the precision and accuracy are very high, around 0.97 and 0.95, and the rates are small (\emph{fpr} is 0.15 and \emph{fnr} is 0.02). In the decoding phase, the results are even more positive, with a precision and accuracy approximately of 0.96 and rates of 0.17 and 0.005 (almost null \emph{fnr} rate). Since there is a trade-off between the two rates, it is interesting to notice that there is a very low \emph{fnr} that leads to a few FPs (wrong alerts). This is advantageous because the alternative would mean missing vulnerabilities. So, these results provide promising evidence of the excellent performance of DEKANT, something that we will be check more thoroughly in the next section.

\begin{table}[h]\scriptsize
	\caption{\small Confusion matrix. \emph{Observed} is the reality, where there are 414 slices with flaws and 96 correct. \emph{Predicted} is the output of DEKANT with our corpus (419 vuln., 91 not vuln. in the evaluation phase; 428 vuln., 82 not vuln. in the decoding phase).}
	\label{t:corpus_cross_val}
	\centering 
	\begin{tabular}{|l|c|c|c|c|c|}							
		
		\multicolumn{2}{c}{}	&	\multicolumn{4}{c}{\textbf{Observed}}	\\ \cline{3-6}
		\multicolumn{2}{c|}{}	&	\multicolumn{2}{c|}{\textbf{Evaluation}}	&	\multicolumn{2}{c|}{\textbf{Decoding}} \\ \cline{3-6}
		\multicolumn{2}{c|}{}	&	Vul	&	N-Vul	&	Vul	&	N-Vul \\ \cline{2-6}
		\multicolumn{1}{c|}{\multirow{2}{*}{\textbf{Predicted}}}	&	Vul	&	405	&	14
		&	412	&	16 \\
		\multicolumn{1}{c|}{}	&	N-Vul	&	9	&	82	&	2	&	80\\ \cline{2-6}
		
	\end{tabular} 
\end{table}


\section{Experimental Evaluation}
\label{s:exper}

Our experimental evaluation addresses the following questions about DEKANT:
(1) Is the tool able to discover novel vulnerabilities? (Section \ref{sb:exper-large});
(2) Can it classify correctly various classes of vulnerabilities? (Section \ref{sb:exper-large});
(3) Is DEKANT more accurate and precise than tools that search for vulnerabilities in plugins (Section \ref{sb:exper-plugins}); tools that do data mining using standard classifiers (Section \ref{sb:exper-dm}); and, tools that do taint analysis (Section \ref{sb:exper-ta})?

\subsection{Open Source Software Evaluation}
\label{sb:exper-large}

This section assesses the ability of DEKANT to classify different vulnerabilities by analyzing 23 WordPress plugins~\cite{Wordpress:15} and 23 packages of real web applications. All of these are written in the PHP language. The plugins are used to determine if the tool is useful for the discovery of new (zero-day) vulnerabilities. The applications serve as a ground truth for the evaluation, since they have known vulnerabilities --- 13 of the packages contain bugs found by~\cite{Medeiros:16} and the other 10 packages have flaws disclosed by various researchers in the past. In every test, DEKANT resorted to the corpus explained in the previous section (however, \emph{none} of the programs utilized in the evaluation was employed to build the corpus). All outputs of the tool were confirmed by us manually to pinpoint valid detections and mistakes.

\begin{table}[b]\scriptsize
	\caption{\small Vulnerable slices in plugins found by DEKANT.}
	\label{t:cms_plugins} 
	\centering 	\begin{tabular}{|l|c|c|c|c|c|c|c|c|}																			
		\hline																			
		\multirow{2}{*}{\textbf{WordPress Plugin}}	&	\multirow{2}{*}{\textbf{Slices}} & \multicolumn{6}{c|}{\textbf{Real Vulnerabilities}}	&		\multirow{2}{*}{\textbf{FP}}	\\ \cline{3-8}										
		&	&	\textbf{SQLI}	&	\textbf{XSS}	&	\textbf{Files\textsuperscript{*}}	&	\textbf{SCD}	&	\textbf{HI}	&	\textbf{CS}	&		\\
		\hline \hline																			
		Appointment Booking Calendar	&	15	&	3	&	4	&		&		&		&		&			\\
		Login by Auth0	&	1	&		&	1	&		&		&		&		&				\\
		Authorizer	&	2	&		&	2	&		&		&		&		&				\\
		BuddyPress	&	4	&		&		&		&		&		&		&			\\
		Contact formgenerator	&	14	&	11	&		&		&		&		&		&			\\
		CP Appointment Calendar	&	11	&	2	&		&		&		&		&		&			\\
		Easy2map	&	13	&		&	1	&	2	&		&		&		&			\\
		Ecwid Shopping Cart	&	1	&		&	1	&		&		&		&		&				\\
		Gantry Framework	&	3	&		&	3	&		&		&		&		&				\\
		Google Maps Travel Route	&	10	&	1	&	2	&		&		&		&		&		1	\\
		Lightbox Plus Colorbox	&	8	&		&	8	&		&		&		&		&				\\
		Payment form for Paypal pro	&	19	&		&	2	&		&		&		&		&			\\
		Recipes writer	&	8	&		&	4	&		&		&		&		&			\\
		ResAds	&	17	&		&	17	&		&		&		&		&				\\
		Simple support ticket system	&	37	&	18	&		&		&		&		&		&			\\
		The Cart Press eCommerce Shopping	&	25	&	8	&	17	&		&		&		&		&				\\
		WebKite	&	1	&	1	&		&		&		&		&		&				\\
		WP Easy Cart eCommerce Shopping	&	78	&	13	&	6	&	29	&	5	&	5	&	2	&			\\
		WP Marketplace	&	45	&	2	&	24	&		&		&		&		&		3	\\
		WP Shop	&	22	&	7	&	10	&		&		&		&		&			\\
		WP ToolBar Removal Node	&	1	&		&	1	&		&		&		&		&		\\
		WP ultimate recipe	&	7	&		&		&		&		&		&		&		1	\\
		WP Web Scraper	&	3	&		&	3	&		&		&		&		&				\\

		\hline\hline																			
		\multicolumn{1}{|r|}{\textbf{Total}}	&	345	&	66	&	106	&	31	&	5	&	5	&	2	&	5	\\
		
		\hline
		\multicolumn{9}{l}{\textsuperscript{*}\ssmall DT \& RFI, LFI vulnerabilities}	
		
	\end{tabular} 
\end{table}

\begin{table*}[t]\scriptsize
	\caption{\small Slices in open source applications processed by DEKANT.}			
	\label{t:WAP-hmm_ta}
	\centering 
	\addtolength{\tabcolsep}{-1.55mm}
	\begin{tabular}{|l|*{20}{c|}}
		\hline
		
		\multirow{2}{*}{\textbf{Web application}}	&	\multirow{2}{*}{\textbf{Version}}	&	\multirow{2}{*}{\textbf{Files}}	&	\multirow{2}{*}{\textbf{LoC}}	&	\textbf{Analysis}	&	\multicolumn{4}{c|}{\textbf{Slices}}							&	\multicolumn{4}{c|}{\textbf{Classification}}							&	\multicolumn{8}{c|}{\textbf{Vulnerability class}}												\\ \cline{6-21}			
		&		&		&		&	\textbf{time (s)}	&	\textbf{Vul}	&	\textbf{San}	&	\textbf{VC}	&	\textbf{Total}	&	\textbf{Vul}	&	\textbf{N-Vul}	&	\textbf{FP}	&	\textbf{FN} 	&	\textbf{SQLI}	&	\textbf{XSS}	&	\textbf{Files\textsuperscript{*}}	&	\textbf{SCD}	&	\textbf{HI}	&	\textbf{CS}	&	\textbf{LDAP}	&	\textbf{SF}	\\
		
		\hline\hline
		
		Admin Control Panel Lite 2	&	0.10.2	&	14	&	1984	&	1	&	81	&	 	&	1	&	82	&	81	&	 	&	1	&	 	&	9	&	72	&		&		&		&		&		&		\\
		Clip Bucket	&	2.7.0.4	&	597	&	148129	&	11	&	22	&	4	&	5	&	31	&	22	&	6	&	3	&	 	&		&	10	&	11	&		&	1	&		&		&		\\
		Clip Bucket	&	2.8	&	606	&	149830	&	12	&	26	&	4	&	5	&	35	&	26	&	6	&	3	&	 	&	4	&	10	&	11	&		&	1	&		&		&		\\
		Ldap address book	&	0.22	&	18	&	4615	&	2	&	40	&	50	&	 	&	90	&	40	&	50	&	 	&	 	&		&	39	&		&		&		&		&	1	&		\\
		Minutes	&	0.42	&	19	&	2670	&	1	&	10	&	 	&	 	&	10	&	10	&	 	&	 	&	 	&	9	&		&		&		&	1	&		&		&		\\
		Mle Moodle	&	0.8.8.5	&	235	&	59723	&	18	&	7	&	 	&	3	&	10	&	6	&	3	&	 	&	1	&		&	5	&	1	&		&		&		&		&		\\
		Php Open Chat	&	3.0.2	&	249	&	83899	&	7	&	11	&	 	&	 	&	11	&	11	&	 	&	 	&	 	&		&	10	&		&		&		&		&	1	&		\\
		Pivotx	&	2.3.10	&	254	&	108893	&	10	&	4	&	3	&	6	&	13	&	4	&	9	&	 	&	 	&		&	1	&	2	&		&		&		&		&	1	\\
		Play sms	&	1.3.1	&	1420	&	248875	&	19	&	6	&	 	&	2	&	8	&	5	&	2	&	 	&	1	&		&	5	&		&		&		&		&		&		\\
		RCR AEsir	&	0.11a	&	8	&	396	&	1	&	13	&	 	&	1	&	14	&	13	&	1	&	 	&	 	&		&	9	&	3	&		&		&	1	&		&		\\
		SAE	&	1.1	&	150	&	47207	&	7	&	148	&	38	&	15	&	201	&	148	&	48	&	5	&	 	&	61	&	65	&	20	&	1	&	1	&		&		&		\\
		Tomahawk Mail	&	2.0	&	155	&	16742	&	3	&	3	&	 	&	3	&	6	&	3	&	3	&	 	&	 	&	2	&	1	&		&		&		&		&		&		\\
		vfront	&	0.99.3	&	438	&	93042	&	15	&	136	&	50	&	30	&	216	&	134	&	78	&	2	&	2	&	32	&	68	&	24	&		&	10	&		&		&		\\

		\hline\hline
		\multicolumn{2}{|c|}{\textbf{Total}}			&	4163	&	966005	&	107	&	507	&	149	&	71	&	727	&	503	&	206	&	14	&	4	&	117	&	295	&	72	&	1	&	14	&	1	&	2	&	1	\\

		\hline
		\multicolumn{21}{l}{\textsuperscript{*}\ssmall DT \& RFI, LFI vulnerabilities}	\\
	\end{tabular}
\end{table*}

\subsubsection{Zero-day Vulnerabilities in Plugins}
\label{sb:exper-extended}

WordPress is the most adopted Content Management System (CMS) worldwide, and therefore its plugins are interesting targets for our study. We selected a diverse set of plugins based on two criteria, the development team and the number of downloads. For the former, we chose 13 plugins built by companies and the other 10 by individual developers. For the second, we picked 10 with less than 20,000 downloads and the other 13 with more than 20,000 downloads. Note that plugins with less downloads were not always those created by individual developers. The plugins were chosen to have also diverse characteristics with regard to the number of files and lines of code (LoC). Although plugins are often believed to be small, in  a few cases they had more than 200 files and 100,000 LoC (see Table \ref{t:cms_plugins_3tools}).

WordPress offers a set of functions to sanitize and validate the data types, to read entry points, and to handle SQL commands (\emph{\$wpdb} class), which are invoked by some of the plugins. Therefore, we configured DEKANT with information about these functions, mapping them to the ISL tokens. Recall that ISL abstracts the PHP instructions, enabling certain behaviors to be captured like sanitization. 

DEKANT extracts 345 slices from the plugins that begin at an entry point and end at a sensitive sink. Next, it translates them into ISL and executes the detection procedure. In total 220 slices are reported as potentially being vulnerable, but 5 of them are actually invalid alarms (i.e., \emph{false positives (FP)}). There are 62 new vulnerabilities that no one had previously found, and 153 bugs that had already been published by other researchers  (Table \ref{t:cms_plugins}).  The remaining slices, a group of 125, are correctly perceived as not vulnerable.  The flaws belong to six classes of vulnerability, ranging from SQLI to CS (columns 3-8). 

The zero-day vulnerabilities appear in 21 plugins: 11 developed by companies and 10 by individual programmers; and 11 having more than 20,000 downloads. The most vulnerable plugin is the one that has more files, while the plugins appearing in the next places are smaller, and the largest plugin in terms of LoC has less than 4 identified bugs. These results reveal that, independently of the development teams, number of downloads, files, and LoC, several of the WordPress plugins used in the wild are insecure.

The new flaws were reported to the developers, and in some cases they have already been acknowledged and fixed, resulting in the release of updated versions of the plugins\footnote{For example, plugins  \emph{appointment-booking-calendar 1.1.7}, \emph{easy2map 1.2.9}, \emph{payment-form-for-paypal-pro 1.0.1}, \emph{resads 1.0.1} and \emph{simple-support-ticket-sys\-tem 1.2} were fixed thanks to this work.}. 
Overall, these experiments are encouraging because the approach demonstrated the potential for the discovery of many classes of vulnerabilities in several open-source plugins, some of them with considerable user bases.

\subsubsection{Real Web Applications}
\label{sb:exper-real_webapps}

To determine if DEKANT is effective at classifying the vulnerabilities belonging to the twelve classes under study, we run the tool with 23 well known vulnerable open source software packages divided into two sets. 

The first set is composed of 13 applications with more than 4,000 files and almost 1 million LoC (Table \ref{t:WAP-hmm_ta}). A few of the packages are large, such as \emph{Play sms} and \emph{Clip Bucket}, with approximately 250 and 150 thousand LoC. There are 727 slices evaluated in this experiment, which were classified manually to enable the validation of the outcomes of DEKANT. Table \ref{t:WAP-hmm_ta}, in columns 6-9, displays the results of this effort, where \emph{Vul} stands for vulnerable slices, \emph{San} for sanitized, and \emph{VC} for validated and/or changed.

DEKANT takes a short time to perform the analysis, in the order of tens of seconds (column 5). Columns 10-13 show that the tool correctly classifies 503 slices as being  vulnerable (\emph{Vul}), 14 slices are wrongly labeled as having bugs (FPs) and 4 have errors that remain undetected (i.e., \emph{false negatives (FN)}). Columns 14-21 present how the 503 slices are sorted out into the twelve classes of vulnerabilities (column Files aggregates three classes). Misclassification (FPs and FNs) is mainly explained by the presence of validation and string modification functions with context-sensitive states. In particular, most FPs belong to the class PHPCI, a type of vulnerability related to the execution of \emph{preg\_match} and \emph{preg\_replace} functions (the remaining were in classes HI and XSS). The FNs are also associated with PHPCI bugs. 

Summing-up, the results are reassuring as DEKANT correctly classifies every vulnerability that was described in \cite{Medeiros:16}, but actually with less FP. The accuracy and precision are very high, around 0.97, and the FP rate is 0.06 and the FN rate is  0.01.

\begin{table}[b]\scriptsize
	\caption{\small Slices in open source software with vulnerabilities disclosed in the past, analyzed by DEKANT.}
	\label{t:total_Vulns}
	\centering 
	\begin{tabular}{|l|*{9}{c|}}
		\hline	
		\multirow{2}{*}{\textbf{Web application}}	&	\multirow{2}{*}{\textbf{Files}}	&	\multirow{2}{*}{\textbf{LoC}}	&	\textbf{Time}	 & \multicolumn{2}{c|}{\textbf{Classif.}}	&	\multicolumn{4}{c|}{\textbf{Vulnerability classes}}			\\ \cline{5-10}
		&		&		&	\textbf{(s)}	&   \textbf{Vul}	&	\textbf{FP}	&	\textbf{SQLI}	&	\textbf{Files\textsuperscript{*}}	&	\textbf{CI\textsuperscript{**}}	&	\textbf{XSS}	\\
		\hline\hline
		cacti-0.8.8b    	&	249	&	95274	&	7	&   2	&	2	&		&		&	1	&	1			\\
		communityEdition	&	228	&	217195	&	21	&   16	&		&	4	&	4	&	3	&	5			\\
		epesi-1.6.0     	&  2246	&	741440	&	90	&   25	&	4	&		&	3	&	 	&	22			\\
		NeoBill0.9-alpha	&	620	&	100139	&	5	&   19	&		&		&	2	&		&	17			\\
		phpMyAdmin-4.2.6	&	538	&	241505	&	12	&   1	&		&		&		&		&	1			\\
		refbase-0.9.6   	&	171	&	109600	&	8	&   5	&	6	&		&		&		&	5	    	\\
		Schoolmate-1.5.4	&	64	&	8411	&	2	&   120 &		&	69	&		&		&	51	    	\\
		VideosTube      	&	39	&	3458	&	2	&   1	&		&		&		&		&	1			\\
		Webchess 1.0    	&	37	&	7704	&	2	&   20	&		&	6	&		&		&	14	    	\\
		Zero-CMS.1.0    	&	21	&	1139	&	2	&   2	&		&	1	&		&		&	1	    	\\

		\hline\hline																					
		\textbf{Total}	&	4213	&	1525865	&	151	&  211	&	12	&	80	&	9	&	4	&	118	    	\\
		
		\hline	
		\multicolumn{3}{l}{\textsuperscript{*}\ssmall DT \& RFI, LFI vulnerabilities}	& \multicolumn{7}{l}{\textsuperscript{**}\ssmall PHPCI vulnerability}\\
	\end{tabular}
\end{table}

For the second set, we run DEKANT with ten applications with flaws previously registered in the  CVE \cite{CVE:14} and NVD \cite{NVD:17} databases (Table \ref{t:total_Vulns}). In total more than 4,200 files and 1.5 million LoC are analyzed. The largest packages are  \emph{epesi} and \emph{phpMyAdmin}, with approximately 750 and 250 thousand LoC. Similarly to the first set of applications, we extracted 310 slices, which were then checked manually.

\begin{table*}[htb]\scriptsize
	\caption{\small Vulnerability discovery results with WordPress plugins for DEKANT, WAPe, and phpSAFE.}
	\label{t:cms_plugins_3tools}
	\centering 											
	\addtolength{\tabcolsep}{-1.15mm}										
	\begin{tabular}{|l|*{18}{c|}}												
		\hline															
		\multirow{2}{*}{\textbf{Plugin}}	&	\multirow{2}{*}{\textbf{Version}}	&	\multirow{2}{*}{\textbf{Files}}	&	\multirow{2}{*}{\textbf{LoC}}	&	\multicolumn{4}{c|}{\textbf{DEKANT}}	&	\multicolumn{5}{c|}{\textbf{WAPe}}	&	\multicolumn{5}{c|}{\textbf{phpSAFE}}	\\ \cline{5-18}									
		&	&		&		&	\textbf{SQLI}	&	\textbf{XSS}	&	\textbf{FP}	&	\textbf{FN}	&	\textbf{SQLI}	&	\textbf{XSS}	&	\textbf{FPP}	&	\textbf{FP}	&	\textbf{FN}	&	\textbf{SQLI}	&	\textbf{XSS}	&	\textbf{FP}	&	\textbf{PFP}	&	\textbf{FN}	\\
		\hline \hline														
		Appointment Booking Calendar	&	1.1.7	&	6	&	2955	&	3	&	4	&		&		&	1	&	3	&	1	&		&	3	&	3	&	4	&	2	&	14	&		\\
		Login by Auth0	&	1.3.6	&	35	&	3101	&		&	1	&		&		&		&	1	&		&		&		&		&	1	&		&		&		\\
		Authorizer	&	2.3.6	&	164	&	159023	&		&	2	&		&		&		&	2	&		&		&		&		&	1	&		&		&	1	\\
		BuddyPress	&	2.4.0	&	574	&	219690	&		&		&		&		&		&		&	1	&		&		&	--	&	--	&	--	&	--	&	--	\\
		Contact formgenerator	&	2.0.1	&	42	&	9187	&	11	&		&		&		&	11	&		&		&		&		&		&		&	3	&		&	11	\\
		CP Appointment Calendar	&	1.1.7	&	7	&	988	&	2	&		&		&		&	2	&		&		&		&		&	2	&		&	9	&		&		\\
		Easy2map	&	1.2.9	&	16	&	3193	&		&	1	&		&		&		&	1	&		&		&		&		&	1	&	8	&	10	&		\\
		Ecwid Shopping Cart	&	3.4.6	&	61	&	16807	&		&	1	&		&		&		&	1	&		&		&		&	--	&	--	&	--	&	--	&	1	\\
		Gantry Framework	&	4.1.6	&	274	&	50717	&		&	3	&		&		&		&	1	&		&		&	2	&		&	1	&		&		&	2	\\
		Google Maps Travel Route	&	1.3.1	&	10	&	1692	&	1	&	2	&	1	&		&	1	&	2	&		&		&		&		&	1	&	7	&	10	&	2	\\
		Lightbox Plus Colorbox	&	2.7.2	&	13	&	5902	&		&	8	&		&		&		&	6	&		&		&	2	&	--	&	--	&	--	&	--	&	8	\\
		Payment form for Paypal pro	&	1.0.1	&	10	&	3920	&		&	2	&		&		&		&	2	&		&		&		&		&	2	&	19	&	2	&		\\
		Recipes writer	&	1.0.4	&	9	&	2074	&		&	4	&		&		&		&	4	&		&		&		&		&	4	&	5	&		&		\\
		ResAds	&	1.0.1	&	30	&	3168	&		&	17	&		&		&		&	2	&		&		&	15	&		&	17	&		&		&		\\
		Simple support ticket system	&	1.2	&	20	&	1533	&	18	&		&		&		&	18	&		&		&		&		&	3	&		&	2	&	7	&	15	\\
		The Cart Press eCommerce Shopping	&	1.4.7	&	220	&	47114	&	8	&	17	&		&		&	8	&	17	&		&		&		&	--	&	--	&	--	&	--	&	25	\\
		WebKite	&	2.0.1	&	13	&	1267	&	1	&		&		&		&	1	&		&		&		&		&		&		&		&		&	1	\\
		WP Easy Cart eCommerce Shopping	&	3.2.3	&	623	&	126448	&	13	&	6	&		&		&	13	&	6	&		&		&		&	--	&	--	&	--	&	--	&	19	\\
		WP Marketplace	&	2.4.1	&	88	&	15485	&	2	&	24	&	3	&	3	&		&	9	&		&	1	&	20	&	2	&	27	&	18	&	30	&		\\
		WP Shop	&	3.5.3	&	49	&	9171	&	7	&	10	&		&		&		&	5	&	1	&		&	12	&	7	&	10	&	5	&	29	&		\\
		WP ToolBar Removal Node	&	1839	&	2	&	544	&		&	1	&		&		&		&	1	&		&		&		&		&	1	&		&		&		\\
		WP ultimate recipe	&	2.5	&	284	&	42774	&		&		&	1	&	1	&		&	1	&		&	1	&		&		&		&	6	&		&	1	\\
		WP Web Scraper	&	3.5	&	89	&	8116	&		&	3	&		&		&		&	3	&		&		&		&	--	&	--	&	--	&	--	&	3	\\

		\hline\hline
		\multicolumn{2}{|r|}{\textbf{Total}}	&	2639	&	734869	&	66	&	106	&	5	&	4	&	55	&	67	&	3	&	2	&	54	&	17	&	70	&	84	&	102	&	89	\\
		
		\hline
		
	\end{tabular}												
\end{table*}

DEKANT classifies 223 slices as having bugs but 12 alarms are invalid (columns 5-6). The vulnerabilities pertain to six classes, where the most common are SQLI and XSS (columns 7-10, with Files aggregating DT, RFI and LFI). The FPs occur in the XSS and PHPCI classes due to equivalent reasons as above. The remaining 87 slices are correctly set as not-vulnerable (not shown in the table). Consequently, we could not find missed bugs (i.e., FN is zero).

Overall, DEKANT had accuracy and precision of 0.96 and 0.95, and a FP rate of 0.12 (and no FNs). These results are very similar to the ones of the first set, demonstrating that the tool is capable of detecting vulnerabilities and of classifying them correctly independently of their classes.

\subsection{Comparison with Plugin Analysis Tools}
\label{sb:exper-plugins}

The section tests plugin analysis tools, namely WAPe~\cite{Medeiros:16} and phpSAFE~\cite{Nunes:15}, and compares them to DEKANT. The two tools implement taint analysis in a diverse manner, but still with the aim of tracking data that flows from the entry points to the sensitive sinks. WAPe is an extension of WAP, and since it is highly configurable, we could set it up with the same knowledge about WordPress functions as DEKANT. phpSAFE only looks for SQLI and XSS vulnerabilities in WordPress plugins. Therefore, to make the comparison among tools fair, we decided to consider only these two classes in the evaluation, and accounted the slices with other bugs as not vulnerable. 
The experiments are based on the 23 plugins previously presented, which have a total of 349 slices (the 345 slices of Section \ref{sb:exper-extended} plus 4 extra slices that were extracted by the other two tools). The results are summarized in Table~\ref{t:cms_plugins_3tools}. 

DEKANT evaluates 345 slices (columns 5-8) and outputs 177 of them as potentially vulnerable to SQLi and XSS. Out of this group, 172 of them have real bugs and 5 are FPs. The remaining 168 slices are correctly classified as not vulnerable. While processing the results, we observed that: (i) there are four vulnerabilities that only DEKANT is able to find; (ii) a few slices with bugs are not collected by DEKANT, which inevitably leads to FNs. This last observation confirms the fundamental role of the slice extractor in these tools, as it gets the paths in code that end up being inspected.

WAPe discovers 122 bugs but misses 54 (columns 9 to 13). The tool includes a false positive predictor, whose aim is to look at the results of taint analysis and exclude bug reports that are potentially invalid --- these are called \emph{false positives predicted (FPP)}. After analysis, three cases are deemed FPP, leaving only two FPs. In the case of DEKANT, these five slices are placed in the non-vulnerable set. WAPe and DEKANT extract 126 slices in common, but there is one slice that is only obtained by the former tool. This slice is correctly classified as vulnerable by WAPe (and causes a FN in the other tools). 

phpSAFE could only process 17 plugins (out of 23) and three of them partially (columns 14 to 18). For this reason, only 234 slices out of 349 are examined. Within the group of analyzed slices, there are 87 vulnerabilities that are found and 33 that are missed. However, phpSAFE finds three errors that no other tool is able to discover. The 84 FPs are caused by the inclusion of sanitization and input change functions in the slices, such as \emph{substr} and \emph{preg\_replace} from PHP and \emph{esc\_attr} and \emph{prepare} from WordPress (the last one protects a SQL statement from SQLI attacks, providing similar functionality as prepared statements).

phpSAFE scans 102 extra slices (aside from the 349 group), which are labeled as \emph{possible false positives (PFP)} in our evaluation. These cases are associated with parts of the code where the results of SQL queries are used in some sink (e.g., to embed database content in a web page returned to a browser). The tool considers any of these results as malicious input, independently of the type of query (e.g., an INSERT or UPDATE SQL command) and the sanitization of query' parameters. In addition, the tool does not seem to correlate these queries with the ones that insert data in the database, and therefore it is difficult to conclude that these slices have any real problem. Therefore, due to this ambiguity, we keep these slices separate from the rest.

\begin{table}[h]\scriptsize
	\caption{\small Evaluation metrics of DEKANT, WAPe, phpSAFE, PhpMinerII, RIPS, and Pixy for the detection of SQLI and XSS.}
	\label{t:total_metrics}	
	\vspace{-3mm}
	\centering 
	\begin{tabular}{|l|*{9}{c|}}
		\hline
		\multirow{2}{*}{\textbf{Metric}}	&	\multicolumn{3}{c|}{\textbf{Plugins}}	&	\multicolumn{3}{c|}{\textbf{WebApps -- Data mining}}	&	\multicolumn{3}{c|}{\textbf{WebApps -- Taint analysis}}	\\	 \cline{2-10}									
		&	\textbf{DEKANT}	&	\textbf{WAPe}	&	\textbf{phpSAFE}	&	\textbf{DEKANT}	&	\textbf{WAPe}	&	\textbf{PhpMiner II}	&	\textbf{DEKANT}	&	\textbf{RIPS}	&	\textbf{Pixy}	\\
		\hline \hline
		acc	&	0.97	&	0.84	&	0.50	&	0.97	&	0.96	&	0.83	&	0.97	&	0.80	&	0.54	\\
		pr	&	0.97	&	0.98	&	0.51	&	0.98	&	0.96	&	0.57	&	0.98	&	0.43	&	0.23	\\
		fpr	&	0.03	&	0.01	&	0.49	&	0.004	&	0.01	&	0.04	&	0.004	&	0.09	&	0.48	\\
		fnr	&	0.02	&	0.31	&	0.51	&	0.14	&	0.15	&	0.74	&	0.14	&	0.69	&	0.37	\\
		\hline
		
		\multicolumn{10}{l}{\ssmall acc: accuracy; pr: precision; fpr: false positive rate; fnr: false negative rate} \\
		
	\end{tabular} \vspace{-2mm}
\end{table}

Table \ref{t:total_metrics} has the metrics results for the three tools (columns 2-4). DEKANT is superior with the highest combined accuracy and precision and low FP and FN rates. WAPe is second, being the tool with the lowest FP rate and the second highest FN rate. phpSAFE has the worst performance, with significantly lower accuracy and precision. Notice that the 102 PFPs of phpSAFE are disregarded from the calculations.

\begin{table*}[!htb]\scriptsize
	\caption{\small Comparison of results between DEKANT, WAPe, PhpMinerII, RIPS and Pixy with open source projects.}	
	\vspace{-3mm}
	\label{t:DEKANT_allTools}
	\centering 
	\addtolength{\tabcolsep}{-1.55mm}
	\begin{tabular}{|l|*{24}{c|}}
		\hline
		\multirow{2}{*}{\textbf{Web application}}	&	\multicolumn{5}{c|}{\textbf{DEKANT}}	&	\multicolumn{6}{c|}{\textbf{WAPe}}	&	\multicolumn{4}{c|}{\textbf{PhpMinerII}}	&	\multicolumn{5}{c|}{\textbf{RIPS}}	&	\multicolumn{4}{c|}{\textbf{Pixy}}	\\ \cline{2-25}
		&	\textbf{SQLI}	&	\textbf{XSS}	&	\textbf{oth}	&	\textbf{FP}	&	\textbf{FN}	&	\textbf{SQLI}	&	\textbf{XSS}	&	\textbf{oth}	&	\textbf{FPP}	&	\textbf{FP}	&	\textbf{FN}	&	\textbf{SQLI}	&	\textbf{XSS}	&	\textbf{FP}	&	\textbf{FN}		&	\textbf{SQLI}	&	\textbf{XSS}	&	\textbf{oth}	&	\textbf{FP}	&	\textbf{FN}	&	\textbf{SQLI}	&	\textbf{XSS}	&	\textbf{FP}	&	\textbf{FN}	\\
		
		\hline\hline
		Admin Control Panel Lite 2	&	9	&	72	&		&	1	&		&	9	&	72	&		&	8	&	1	&		&	9	&	23	&	1	&	49	&	9	&	7	&		&	7	&	65	&	9	&	67	&	12	&	5	\\
		Clip Bucket	&		&	10	&	12	&	3	&	9	&		&	10	&	12	&	2	&	4	&	9	&		&	19	&	20	&		&	--	&	--	&	--	&	--	&	31	&		&	19	&	47	&		\\
		Clip Bucket	&	4	&	10	&	12	&	3	&	9	&	4	&	10	&	12	&	2	&	4	&	9	&	3	&	19	&	17	&	1	&	--	&	--	&	--	&	--	&	35	&	3	&	19	&	47	&	1	\\
		Ldap address book	&		&	39	&	1	&		&	3	&		&	36	&	1	&		&	2	&	6	&		&		&	&	39	&	&	1	&	4	&	5	&	38	&		&		&		&	39	\\
		Minutes	&	9	&		&	1	&		&	9	&	6	&		&	1	&		&		&	12	&		&	5	&	7	&	11	&	&	7	&	3	&	10	&	9	&		&	7	&	55	&	9	\\
		Mle Moodle	&		&	5	&	1	&		&	14	&		&	5	&	1	&	3	&		&	14	&		&	10	&	27	&	8	&		&	6	&	2	&	7	&	12	&		&	18	&	621	&		\\
		Php Open Chat	&		&	10	&	1	&		&	7	&		&	9	&	1	&		&		&	8	&		&	9	&	7	&	8	&		&	17	&		&	43	&	1	&		&	2	&	26	&	15	\\
		Pivotx	&		&	1	&	3	&		&	10	&		&	1	&	3	&	9	&		&	10	&		&	4	&	1	&	6	&		&	6	&	4	&	7	&	4	&		&	4	&	16	&	6	\\
		Play sms	&		&	5	&		&		&	7	&		&	5	&		&	2	&		&	7	&		&	10	&	12	&		&	&	6	&	2	&	31	&	4	&		&	10	&	20	&		\\
		RCR AEsir	&		&	9	&	4	&		&		&		&	9	&	4	&	1	&		&		&		&	3	&		&	6	&	&	8	&	4	&	2	&	1	&		&	2	&		&	7	\\
		SAE	&	61	&	65	&	22	&	5	&		&	61	&	65	&	20	&		&	10	&	2	&		&	8	&	2	&	118	&		&	2	&	5	&		&	141	&		&	65	&	178	&	61	\\
		Tomahawk Mail	&	2	&	1	&		&		&		&	2	&	1	&		&	3	&		&		&		&	1	&	1	&	2	&		&	2	&		&	6	&	1	&		&	1	&	1	&	2	\\
		vfront	&	32	&	68	&	34	&	2	&	11	&	32	&	68	&	34	&	24	&	2	&	11	&		&	1	&		&	105	&	&	74	&	39	&	114	&	32	&		&	70	&	6	&	36	\\
		
		\hline\hline
		\textbf{Total}	&	117	&	295	&	91	&	14	&	79	&	114	&	291	&	89	&	54	&	23	&	88	&	12	&	112	&	95	&	353	&	9	&	136	&	63	&	232	&	374	&	12	&	289	&	1031	&	181	\\
		\hline
	\end{tabular} \vspace{-2mm}
	\vspace{-1mm}
\end{table*}

\subsection{Comparison with Data Mining Tools}
\label{sb:exper-dm}

A few other tools have implemented data mining mechanisms for tasks related with bug discovery, namely WAPe and PhpMinerII~\cite{Shar:12a,Shar:12b}. WAPe and PhpMinerII classify slices by resorting to data mining with standard classifiers, which do not consider order. WAPe obtains the slices with taint analysis and then predicts if they are FPs or TPs with the classifiers, with the aim of reducing the alerts that are generated by mistake. PhpMinerII uses data mining to find out if slices hold attributes that make them look vulnerable, without specific concerns about false positives. This tool handles only SQLI and reflected XSS vulnerabilities.

Since PhpMinerII is not configurable with information about WordPress, and consequently it would perform much worse with plugins, we opted to experiment with the first set of 13 application packages. Similarly, the same limitation applies to the vulnerability detection tools that will be studied in the next section, and so we will focus on these applications for the rest of the evaluations. 

We observed that the various tools (from this and next section) survey different groups of slices because of their specific implementation of the slice extractor. Therefore, we decided to create a superset with all slices that could be captured based on the outputs of the tools, which contains 2609 slices. This set was then manually examined to determine which slices are vulnerable, and it serves as a ground truth. Overall there are 582 slices with vulnerabilities (117 SQLI, 360 XSS, and 105 others) and 2027 slices without problems. This second group was divided in a few subsets, namely, slices with sanitized input, slices with validated or modified input, and slices without external sources (i.e., without entry points) but with a sensitive sink. This last group was provided by PhpMinerII and we designate it as the \emph{no-source} subset.

\subsubsection{All Vulnerability Classes}
\label{ss:exper-dm-all}

A summary of the experimental results is included in Table~\ref{t:DEKANT_allTools}. The vulnerabilities are distributed by classes SQLI, XSS and others, to facilitate the assessment of alternative tools that only address specific bugs (like PhpMinerII). Columns 2 to 6 are about DEKANT, displaying a total of 503 identified bugs. Notice that there are 75 more FNs than in Table~\ref{t:WAP-hmm_ta} because now we are covering a larger number of slices, some of which are not extracted by DEKANT. The next six columns display WAPe's results. WAPe reports less vulnerabilities and a few more FPs and FNs.

With regard to false positives, DEKANT judges correctly as not vulnerable the 71 validated and/or changed slices  (i.e., column \emph{VC} in Table \ref{t:WAP-hmm_ta}) but  WAPe just predicts 48 of them as FPP. Even though WAPe handles a considerable number of symptoms to reduce mistakes, there is a lack of attribute relation verification that induces erroneous decisions --- the tool only checks if attributes exist in a slice but does not have a way to relate them. 

The difference in false negatives between the tools is also explained by the same reason, plus the importance of considering the order of the code elements in the slice. In particular, a misclassification can occur when there is a concatenation of  tainted with untainted variables  (i.e., which were validated or modified); this causes the data mining classifier to find symptoms related with validation and outputs the slices as FPs. DEKANT implements a sequence model that takes into account how the code elements appear in the slice, prevailing in these situations.

\begin{table}[b]\scriptsize
	\caption{\small Confusion matrix of DEKANT, WAPe and RIPS for the detection of all vulnerability classes.}
	\label{t:TA_Matrix_metrics_all}
	\centering \vspace{-2mm}
	\begin{tabular}{|c|c|c||c|c||c|cc|c|c|c|}							
		
		\cline{8-11}
		\multicolumn{1}{c}{}	&	\multicolumn{6}{c}{\textbf{Observed}}	&	\multicolumn{1}{||c|}{\textbf{Metric}}	&	\textbf{DEKANT}	&	\textbf{WAPe}	&	\textbf{RIPS}	\\ \cline{2-11}
		\multicolumn{1}{c|}{}	&	\multicolumn{2}{c||}{\textbf{DEKANT}}	&\multicolumn{2}{c||}{\textbf{WAPe}}	&\multicolumn{2}{c|}{\textbf{RIPS}}	&	\multicolumn{1}{||l|}{acc}	&	0.96	&	0.96	&	0.77	\\ \cline{2-7}
		\multicolumn{1}{c|}{\textbf{Predicted}}	&	Vul	&	N-Vul	&	Vul	&	N-Vul	&	Vul	&	N-Vul	 &	\multicolumn{1}{||l|}{pr}	&	0.97	&	0.96	&	0.47	\\ \cline{1-7}
		Vul	&	503	&	14  &	494	&	23  &	208	&	232  &	\multicolumn{1}{||l|}{fpr}	&	0.007	&	0.01	& 	0.11	\\
		N-Vul	&	79	&	2013  &	88	&	2004  &	374	&	1795  &	\multicolumn{1}{||l|}{fnr}	&	0.13	&	0.15	& 	0.64	\\ \cline{1-11}
		\multicolumn{11}{l}{\ssmall acc: accuracy; pr: precision; fpr: false positive rate; fnr: false negative rate} \\
	\end{tabular} 
\end{table}

Table \ref{t:TA_Matrix_metrics_all} sums up de evaluation, combining the confusion matrix and metrics. The results are encouraging with DEKANT performing better than WAPe, namely because it shows superior FP and FN rates.

\subsubsection{Just SQLI and XSS}
\label{ss:exper-dm_sqli_xss}

This subsection only considers SQLI and reflected XSS for a fair comparison with PhpMinerII. PhpMinerII does not come trained when downloaded, and so we had to build a dataset for that purpose. 
The training dataset was constructed by recreating the procedure explained in~\cite{Shar:12a,Shar:12b}, where the WEKA package implemented the data mining tasks~\cite{Witten:11}. 
The same classifiers were evaluated to select the best. Overall, the C4.5/J48 classifier was chosen, with an accuracy and precision close to 0.92.

Table \ref{t:DEKANT_allTools} has the results for PhpMinerII. The tool obtains 1052 slices, where 219 are reported  as vulnerable and 833 as not-vulnerable. Manually, we inspected these slices and found out that only 604 were correctly labeled, 124 as vulnerable and 480 as not-vulnerable. Consequently, the tool generates 95 FP and 353 FN. This notable misclassification is explained by various factors, such as missing validations and string modifications of inputs, and  not taking into account the order of code elements. In addition, some of the slides belong to the no-source subset and they lead necessarily to invalid alarms (as there is no entry point to be maliciously exploited). 

DEKANT outputs 412 vulnerabilities and 8 incorrect reports (out of the 14 shown in table). It also misses 65 slices with bugs (out of the 79 shown in table). WAPe classifies 405 vulnerabilities, but with 16 FPs (of the 23 presented in table) and 72 FNs (out of the 88). Only 82 of the 124 identified bugs by PhpMinerII are also flagged as being vulnerable by DEKANT and WAPe. This means that the 42 remaining vulnerable slices justify the increase of FN in the two tools.

Table \ref{t:total_metrics} displays the calculated metrics when only SQLI and XSS are contemplated.
DEKANT and WAPe surpass PhpMinerII, exhibiting higher quality values for all metrics. Both DEKANT and WAPe have an excellent accuracy and precision, but the former is superior with 0.97 and 0.98 on the metrics. In addition, DEKANT has better rates for false positives and false negatives.

\subsection{Comparison with Taint Analysis Tools}
\label{sb:exper-ta}

There have been tools proposed in the past that perform taint analysis to locate vulnerabilities, and two notable examples are RIPS~\cite{Dahse:14} and Pixy~\cite{Jovanovic:06}. They track data arriving at the entry points to determine if it reaches a sensitive sink, taking sanitization operations in consideration. RIPS detects the same classes of vulnerabilities as DEKANT, but Pixy only looks for SQLI and reflected XSS. Our evaluation compares the three tools while processing the same applications of the previous section (i.e., the dataset with 2609 slices), with results being displayed in Table \ref{t:DEKANT_allTools}.

\subsubsection{All Vulnerability Classes}

The RIPS tool only outputs information about slices that are regarded as vulnerable. Therefore, when no result appears for a particular slice, this could occur because the slice was considered valid or due to the inability to extract the slice. Since we are unable to separate the two situations, this brings some level of uncertainty to the analysis.

RIPS generates alerts for a total of 440 slices in 11 applications (of the 13). Out of this group, 208 correspond to slices with real bugs and the remaining 232 to false alerts. These FPs occur essentially in slices with functions that change the data received at the entry points (such as, \emph{substr} and \emph{preg\_replace}) or in slices with validation functions. This demonstrates the importance of the identification of false positive symptoms and of evaluating the slices taking into consideration the order of code elements (like DEKANT does). RIPS does not catch 374 vulnerabilities from the ground truth. We speculate that the reason for this high number of FNs is probably related to the extractor being unable to gather many slices.

Table \ref{t:TA_Matrix_metrics_all} has the confusion matrix and the metrics. RIPS is outperformed by both DEKANT and WAPe in the dataset with all classes of flaws. Its accuracy and precision are 0.77 and 0.47, which are not as high as the other tools.

\subsubsection{Just SQLI and XSS}

Pixy only searches for SQLI and XSS vulnerabilities. Therefore, our evaluation just covers these two classes (meaning that the 105 slices with \emph{other} vulnerabilities are treated as being true negatives).

Pixy results are displayed in the last four columns of  Table \ref{t:DEKANT_allTools}. The tool raises 1332  alerts, but the majority of them are mistakes\footnote{In fact, a significant number of non-vulnerable slices included in our ground truth dataset comes from Pixy.}. Only 301 reported vulnerabilities are real (12 SQLI and 289 XSS). There are 176 undetected bugs. Curiously, Pixy has around half the FNs of RIPS, and for some applications it is the tool that detects more XSS vulnerabilities (e.g., the \emph{Minutes} application) but at the cost of a high FP rate.
Section \ref{ss:exper-dm_sqli_xss} presents the details about the DEKANT evaluation for SQLI and XSS bugs. In what concerns RIPS, the tool finds 145 buggy slices but misses 328 (out of the 374 in the table). It also wrongly reports 192 slices as being vulnerable (of the 232 in the table).

Table \ref{t:total_metrics} presents the metrics for these tools.
The results corroborate the promising detection capabilities of DEKANT, as the tool has the best accuracy and precision and the lowest FP and FN rates. RIPS is second, but with an accuracy and precision reasonably below. For our dataset, the weakest values are obtained by Pixy, with a small precision due to the many FP.

\section{Conclusion}
\label{conclusion}

The paper explores a new approach to detect web application vulnerabilities inspired in NLP in which static analysis tools \emph{learn} to detect vulnerabilities automatically using machine learning. Whereas in classical static analysis tools it is necessary to code knowledge about how each vulnerability is detected, our approach obtains knowledge about vulnerabilities automatically.
The approach uses a sequence model (HMM) that, first, learns to characterize vulnerabilities from a corpus composed of sequences of observations annotated as vulnerable or not, then processes new sequences of observations based on this knowledge, taking into consideration the order in which the observations appear.
The model can be used as a static analysis tool to discover vulnerabilities in source code and identify their location.

\begin{acks}
This work was partially supported by the national funds through Funda\c{c}\~ao para a Ci\^encia e a Tecnologia (FCT)/MCTES (PIDDAC)/FEDER with reference to project AAC-2/SAICT/2017-029058 (SEAL), and through FCT with references UID/CEC/00408/2019 (LASIGE) and UID/CEC/50021/2019 (INESC-ID).
\end{acks}

\bibliographystyle{ACM-Reference-Format}
\bibliography{ref-sec-groups}


\begin{thebibliography}{44}


\ifx \showCODEN    \undefined \def \showCODEN     #1{\unskip}     \fi
\ifx \showDOI      \undefined \def \showDOI       #1{#1}\fi
\ifx \showISBNx    \undefined \def \showISBNx     #1{\unskip}     \fi
\ifx \showISBNxiii \undefined \def \showISBNxiii  #1{\unskip}     \fi
\ifx \showISSN     \undefined \def \showISSN      #1{\unskip}     \fi
\ifx \showLCCN     \undefined \def \showLCCN      #1{\unskip}     \fi
\ifx \shownote     \undefined \def \shownote      #1{#1}          \fi
\ifx \showarticletitle \undefined \def \showarticletitle #1{#1}   \fi
\ifx \showURL      \undefined \def \showURL       {\relax}        \fi
\providecommand\bibfield[2]{#2}
\providecommand\bibinfo[2]{#2}
\providecommand\natexlab[1]{#1}
\providecommand\showeprint[2][]{arXiv:#2}

\bibitem[\protect\citeauthoryear{Arisholm, Briand, and Johannessen}{Arisholm
  et~al\mbox{.}}{2010}]%
        {Arisholm:10}
\bibfield{author}{\bibinfo{person}{Erik Arisholm}, \bibinfo{person}{Lionel~C
  Briand}, {and} \bibinfo{person}{Eivind~B Johannessen}.}
  \bibinfo{year}{2010}\natexlab{}.
\newblock \showarticletitle{A systematic and comprehensive investigation of
  methods to build and evaluate fault prediction models}.
\newblock \bibinfo{journal}{\emph{Journal of Systems and Software}}
  \bibinfo{volume}{83}, \bibinfo{number}{1} (\bibinfo{year}{2010}),
  \bibinfo{pages}{2--17}.
\newblock


\bibitem[\protect\citeauthoryear{Backes, Rieck, Skoruppa, Stock, and
  Yamaguchi}{Backes et~al\mbox{.}}{2017}]%
        {Backes:17}
\bibfield{author}{\bibinfo{person}{Michael Backes}, \bibinfo{person}{Konrad
  Rieck}, \bibinfo{person}{Malte Skoruppa}, \bibinfo{person}{Ben Stock}, {and}
  \bibinfo{person}{Fabian Yamaguchi}.} \bibinfo{year}{2017}\natexlab{}.
\newblock \showarticletitle{Efficient and Flexible Discovery of {PHP}
  Application Vulnerabilities}. In \bibinfo{booktitle}{\emph{EuroS{\&}P}}.
  \bibinfo{pages}{334--349}.
\newblock


\bibitem[\protect\citeauthoryear{Baum and Petrie}{Baum and Petrie}{1966}]%
        {Baum:66}
\bibfield{author}{\bibinfo{person}{Leonard~E. Baum} {and} \bibinfo{person}{Ted
  Petrie}.} \bibinfo{year}{1966}\natexlab{}.
\newblock \showarticletitle{Statistical Inference for Probabilistic Functions
  of Finite State Markov Chains}.
\newblock \bibinfo{journal}{\emph{The Annals of Mathematical Statistics}}
  \bibinfo{volume}{37}, \bibinfo{number}{6} (\bibinfo{year}{1966}),
  \bibinfo{pages}{1554--1563}.
\newblock


\bibitem[\protect\citeauthoryear{{BBC Technology}}{{BBC Technology}}{2014}]%
        {BBC:14}
\bibfield{author}{\bibinfo{person}{{BBC Technology}}.}
  \bibinfo{year}{2014}\natexlab{}.
\newblock \bibinfo{title}{Millions of websites hit by {Drupal} hack attack}.
\newblock
\newblock
\newblock
\shownote{http://www.bbc.com/news/technology-29846539.}


\bibitem[\protect\citeauthoryear{CVE}{CVE}{[n. d.]}]%
        {CVE:14}
\bibfield{author}{\bibinfo{person}{CVE}.} \bibinfo{year}{[n. d.]}\natexlab{}.
\newblock
\newblock
\newblock
\shownote{{h}ttp://cve.mitre.org.}


\bibitem[\protect\citeauthoryear{Dahse and Holz}{Dahse and Holz}{2014}]%
        {Dahse:14}
\bibfield{author}{\bibinfo{person}{Johannes Dahse} {and}
  \bibinfo{person}{Thorsten Holz}.} \bibinfo{year}{2014}\natexlab{}.
\newblock \showarticletitle{Simulation of Built-in {PHP} Features for Precise
  Static Code Analysis}. In \bibinfo{booktitle}{\emph{Proceedings of the 21st
  Network and Distributed System Security Symposium}}.
\newblock


\bibitem[\protect\citeauthoryear{Dahse and Holz}{Dahse and Holz}{2015}]%
        {Dahse:15}
\bibfield{author}{\bibinfo{person}{Johannes Dahse} {and}
  \bibinfo{person}{Thorsten Holz}.} \bibinfo{year}{2015}\natexlab{}.
\newblock \showarticletitle{Experience Report: An Empirical Study of {PHP}
  Security Mechanism Usage}. In \bibinfo{booktitle}{\emph{Proceedings of the
  2015 International Symposium on Software Testing and Analysis}}.
  \bibinfo{pages}{60--70}.
\newblock


\bibitem[\protect\citeauthoryear{Dem\v{s}ar}{Dem\v{s}ar}{2006}]%
        {Demsar:06}
\bibfield{author}{\bibinfo{person}{Janez Dem\v{s}ar}.}
  \bibinfo{year}{2006}\natexlab{}.
\newblock \showarticletitle{Statistical Comparisons of Classifiers over
  Multiple Data Sets}.
\newblock \bibinfo{journal}{\emph{The Journal of Machine Learning Research}}
  \bibinfo{volume}{7} (\bibinfo{date}{Dec} \bibinfo{year}{2006}),
  \bibinfo{pages}{1--30}.
\newblock


\bibitem[\protect\citeauthoryear{Fonseca and Vieira}{Fonseca and
  Vieira}{2014}]%
        {Fonseca:14}
\bibfield{author}{\bibinfo{person}{Jos\'{e} Fonseca} {and}
  \bibinfo{person}{Marco Vieira}.} \bibinfo{year}{2014}\natexlab{}.
\newblock \showarticletitle{A Practical Experience on the Impact of Plugins in
  Web Security}. In \bibinfo{booktitle}{\emph{Proceedings of the 33rd IEEE
  Symposium on Reliable Distributed Systems}}. \bibinfo{pages}{21--30}.
\newblock


\bibitem[\protect\citeauthoryear{{{HELPNETSECURITY}}}{{{HELPNETSECURITY}}}{2017}]%
        {HelpNetSecurity:17}
\bibfield{author}{\bibinfo{person}{{{HELPNETSECURITY}}}.}
  \bibinfo{year}{2017}\natexlab{}.
\newblock \bibinfo{title}{Hacker breached 60+ unis, govt agencies via {SQL}
  injection}.
\newblock
\newblock
\newblock
\shownote{https://www.helpnetsecurity.com/2017/02/16/hacker-govt-agencies-via-sql-injection/.}


\bibitem[\protect\citeauthoryear{Imperva}{Imperva}{2017}]%
        {Imperva:17}
\bibfield{author}{\bibinfo{person}{Imperva}.} \bibinfo{year}{2017}\natexlab{}.
\newblock \bibinfo{title}{The State of Web Application Vulnerabilities in
  2017}.  (\bibinfo{date}{Dec.} \bibinfo{year}{2017}).
\newblock


\bibitem[\protect\citeauthoryear{Jovanovic, Kruegel, and Kirda}{Jovanovic
  et~al\mbox{.}}{2006}]%
        {Jovanovic:06}
\bibfield{author}{\bibinfo{person}{N. Jovanovic}, \bibinfo{person}{C. Kruegel},
  {and} \bibinfo{person}{E. Kirda}.} \bibinfo{year}{2006}\natexlab{}.
\newblock \showarticletitle{Precise alias analysis for static detection of web
  application vulnerabilities}. In \bibinfo{booktitle}{\emph{Proceedings of the
  2006 Workshop on Programming Languages and Analysis for Security}}.
  \bibinfo{pages}{27--36}.
\newblock


\bibitem[\protect\citeauthoryear{Jurafsky and Martin}{Jurafsky and
  Martin}{2008}]%
        {Jurafsky:08}
\bibfield{author}{\bibinfo{person}{Daniel Jurafsky} {and}
  \bibinfo{person}{James~H. Martin}.} \bibinfo{year}{2008}\natexlab{}.
\newblock \bibinfo{booktitle}{\emph{{Speech and Language Processing}}}.
\newblock \bibinfo{publisher}{Prentice Hall}.
\newblock


\bibitem[\protect\citeauthoryear{Lessmann, Baesens, Mues, and Pietsch}{Lessmann
  et~al\mbox{.}}{2008}]%
        {Lessmann:08}
\bibfield{author}{\bibinfo{person}{Stefan Lessmann}, \bibinfo{person}{Bart
  Baesens}, \bibinfo{person}{Christophe Mues}, {and} \bibinfo{person}{Swantje
  Pietsch}.} \bibinfo{year}{2008}\natexlab{}.
\newblock \showarticletitle{Benchmarking classification models for software
  defect prediction: A proposed framework and novel findings}.
\newblock \bibinfo{journal}{\emph{IEEE Transactions on Software Engineering}}
  \bibinfo{volume}{34}, \bibinfo{number}{4} (\bibinfo{year}{2008}),
  \bibinfo{pages}{485--496}.
\newblock


\bibitem[\protect\citeauthoryear{Li, Zou, Xu, Ou, Jin, Wang, Deng, and
  Zhong}{Li et~al\mbox{.}}{2018}]%
        {Li:18}
\bibfield{author}{\bibinfo{person}{Zhen Li}, \bibinfo{person}{Deqing Zou},
  \bibinfo{person}{Shouhuai Xu}, \bibinfo{person}{Xinyu Ou},
  \bibinfo{person}{Hai Jin}, \bibinfo{person}{Sujuan Wang},
  \bibinfo{person}{Zhijun Deng}, {and} \bibinfo{person}{Yuyi Zhong}.}
  \bibinfo{year}{2018}\natexlab{}.
\newblock \showarticletitle{{VulDeePecker}: {A} Deep Learning-Based System for
  Vulnerability Detection}. In \bibinfo{booktitle}{\emph{Annual Network and
  Distributed System Security Symposium}}.
\newblock


\bibitem[\protect\citeauthoryear{Medeiros, Neves, and Correia}{Medeiros
  et~al\mbox{.}}{2016a}]%
        {Medeiros:16a}
\bibfield{author}{\bibinfo{person}{Ib\'{e}ria Medeiros},
  \bibinfo{person}{Nuno~F. Neves}, {and} \bibinfo{person}{Miguel Correia}.}
  \bibinfo{year}{2016}\natexlab{a}.
\newblock \showarticletitle{{DEKANT}: a static analysis tool that learns to
  detect web application vulnerabilities}. In
  \bibinfo{booktitle}{\emph{Proceedings of the 25th International Symposium on
  Software Testing and Analysis}}.
\newblock


\bibitem[\protect\citeauthoryear{Medeiros, Neves, and Correia}{Medeiros
  et~al\mbox{.}}{2016b}]%
        {Medeiros:15a}
\bibfield{author}{\bibinfo{person}{Ib\'{e}ria Medeiros},
  \bibinfo{person}{Nuno~F. Neves}, {and} \bibinfo{person}{Miguel Correia}.}
  \bibinfo{year}{2016}\natexlab{b}.
\newblock \showarticletitle{Detecting and Removing Web Application
  Vulnerabilities with Static Analysis and Data Mining}.
\newblock \bibinfo{journal}{\emph{IEEE Transactions on Reliability}}
  \bibinfo{volume}{65}, \bibinfo{number}{1} (\bibinfo{date}{March}
  \bibinfo{year}{2016}), \bibinfo{pages}{54--69}.
\newblock


\bibitem[\protect\citeauthoryear{Medeiros, Neves, and Correia}{Medeiros
  et~al\mbox{.}}{2016c}]%
        {Medeiros:16}
\bibfield{author}{\bibinfo{person}{Ib\'{e}ria Medeiros},
  \bibinfo{person}{Nuno~F. Neves}, {and} \bibinfo{person}{Miguel Correia}.}
  \bibinfo{year}{2016}\natexlab{c}.
\newblock \showarticletitle{Equipping {WAP} with Weapons to Detect
  Vulnerabilities}. In \bibinfo{booktitle}{\emph{Proceedings of the 46th Annual
  IEEE/IFIP International Conference on Dependable Systems and Networks}}.
\newblock


\bibitem[\protect\citeauthoryear{Neuhaus, Zimmermann, Holler, and
  Zeller}{Neuhaus et~al\mbox{.}}{2007}]%
        {Neuhaus:07}
\bibfield{author}{\bibinfo{person}{Stephan Neuhaus}, \bibinfo{person}{Thomas
  Zimmermann}, \bibinfo{person}{Christian Holler}, {and}
  \bibinfo{person}{Andreas Zeller}.} \bibinfo{year}{2007}\natexlab{}.
\newblock \showarticletitle{Predicting vulnerable software components}. In
  \bibinfo{booktitle}{\emph{Proceedings of the 14th ACM Conference on Computer
  and Communications Security}}. \bibinfo{pages}{529--540}.
\newblock


\bibitem[\protect\citeauthoryear{Nunes, Fonseca, and Vieira}{Nunes
  et~al\mbox{.}}{2015}]%
        {Nunes:15}
\bibfield{author}{\bibinfo{person}{Paulo Nunes}, \bibinfo{person}{Jos\'{e}
  Fonseca}, {and} \bibinfo{person}{Marco Vieira}.}
  \bibinfo{year}{2015}\natexlab{}.
\newblock \showarticletitle{{phpSAFE}: A Security Analysis Tool for {OOP} Web
  Application Plugins}. In \bibinfo{booktitle}{\emph{Proceedings of the 45th
  Annual IEEE/IFIP International Conference on Dependable Systems and
  Networks}}.
\newblock


\bibitem[\protect\citeauthoryear{NVD}{NVD}{[n. d.]}]%
        {NVD:17}
\bibfield{author}{\bibinfo{person}{NVD}.} \bibinfo{year}{[n. d.]}\natexlab{}.
\newblock
\newblock
\newblock
\shownote{{h}ttp://nvd.nist.org.}


\bibitem[\protect\citeauthoryear{Perl, Dechand, Smith, Arp, Yamaguchi, Rieck,
  Fahl, and Acar}{Perl et~al\mbox{.}}{2015}]%
        {Perl:15}
\bibfield{author}{\bibinfo{person}{Henning Perl}, \bibinfo{person}{Sergej
  Dechand}, \bibinfo{person}{Matthew Smith}, \bibinfo{person}{Daniel Arp},
  \bibinfo{person}{Fabian Yamaguchi}, \bibinfo{person}{Konrad Rieck},
  \bibinfo{person}{Sascha Fahl}, {and} \bibinfo{person}{Yasemin Acar}.}
  \bibinfo{year}{2015}\natexlab{}.
\newblock \showarticletitle{{VCCFinder}: Finding Potential Vulnerabilities in
  Open-Source Projects to Assist Code Audits}. In
  \bibinfo{booktitle}{\emph{Proceedings of the 22nd ACM SIGSAC Conference on
  Computer and Communications Security}} \emph{(\bibinfo{series}{CCS '15})}.
  \bibinfo{pages}{426--437}.
\newblock


\bibitem[\protect\citeauthoryear{Rabiner}{Rabiner}{1989}]%
        {Rabiner:89}
\bibfield{author}{\bibinfo{person}{Lawrence~R Rabiner}.}
  \bibinfo{year}{1989}\natexlab{}.
\newblock \showarticletitle{A tutorial on hidden {Markov} models and selected
  applications in speech recognition}.
\newblock \bibinfo{journal}{\emph{Proc. IEEE}} \bibinfo{volume}{77},
  \bibinfo{number}{2} (\bibinfo{year}{1989}), \bibinfo{pages}{257--286}.
\newblock


\bibitem[\protect\citeauthoryear{Rasthofer, Arzt, and Bodden}{Rasthofer
  et~al\mbox{.}}{2014}]%
        {Rasthofer:14}
\bibfield{author}{\bibinfo{person}{S. Rasthofer}, \bibinfo{person}{S. Arzt},
  {and} \bibinfo{person}{E. Bodden}.} \bibinfo{year}{2014}\natexlab{}.
\newblock \showarticletitle{A Machine-learning Approach for Classifying and
  Categorizing Android Sources and Sinks}. In
  \bibinfo{booktitle}{\emph{Proceedings of the 2014 Network and Distributed
  System Security Symposium (NDSS)}}.
\newblock


\bibitem[\protect\citeauthoryear{Russell, Kim, Hamilton, Lazovich, Harer,
  Ozdemir, Ellingwood, and McConley}{Russell et~al\mbox{.}}{2018}]%
        {Russell:18}
\bibfield{author}{\bibinfo{person}{Rebecca~L. Russell},
  \bibinfo{person}{Louis~Y. Kim}, \bibinfo{person}{Lei~H. Hamilton},
  \bibinfo{person}{Tomo Lazovich}, \bibinfo{person}{Jacob~A. Harer},
  \bibinfo{person}{Onur Ozdemir}, \bibinfo{person}{Paul~M. Ellingwood}, {and}
  \bibinfo{person}{Marc~W. McConley}.} \bibinfo{year}{2018}\natexlab{}.
\newblock \showarticletitle{Automated Vulnerability Detection in Source Code
  Using Deep Representation Learning}. In \bibinfo{booktitle}{\emph{Proceedings
  of the International Conference on Machine Learning and Application
  (ICMLA)}}.
\newblock


\bibitem[\protect\citeauthoryear{Scandariato, Walden, Hovsepyan, and
  Joosen}{Scandariato et~al\mbox{.}}{2014}]%
        {Scandariato:14}
\bibfield{author}{\bibinfo{person}{R. Scandariato}, \bibinfo{person}{J.
  Walden}, \bibinfo{person}{A. Hovsepyan}, {and} \bibinfo{person}{W. Joosen}.}
  \bibinfo{year}{2014}\natexlab{}.
\newblock \showarticletitle{Predicting Vulnerable Software Components via Text
  Mining}.
\newblock \bibinfo{journal}{\emph{IEEE Transactions on Software Engineering}}
  \bibinfo{volume}{40}, \bibinfo{number}{10} (\bibinfo{year}{2014}),
  \bibinfo{pages}{993--1006}.
\newblock


\bibitem[\protect\citeauthoryear{Shankar, Talwar, Foster, and Wagner}{Shankar
  et~al\mbox{.}}{2001}]%
        {Shankar:01}
\bibfield{author}{\bibinfo{person}{Umesh Shankar}, \bibinfo{person}{Kunal
  Talwar}, \bibinfo{person}{Jeffrey~S Foster}, {and} \bibinfo{person}{David
  Wagner}.} \bibinfo{year}{2001}\natexlab{}.
\newblock \showarticletitle{Detecting format-string vulnerabilities with type
  qualifiers}. In \bibinfo{booktitle}{\emph{Proceedings of the 10th USENIX
  Security Symposium}}.
\newblock


\bibitem[\protect\citeauthoryear{Shar and Tan}{Shar and Tan}{2012a}]%
        {Shar:12a}
\bibfield{author}{\bibinfo{person}{Lwin~Khin Shar} {and} \bibinfo{person}{Hee
  Beng~Kuan Tan}.} \bibinfo{year}{2012}\natexlab{a}.
\newblock \showarticletitle{Mining input sanitization patterns for predicting
  {SQL} injection and cross site scripting vulnerabilities}. In
  \bibinfo{booktitle}{\emph{Proceedings of the 34th International Conference on
  Software Engineering}}. \bibinfo{pages}{1293--1296}.
\newblock


\bibitem[\protect\citeauthoryear{Shar and Tan}{Shar and Tan}{2012b}]%
        {Shar:12b}
\bibfield{author}{\bibinfo{person}{Lwin~Khin Shar} {and} \bibinfo{person}{Hee
  Beng~Kuan Tan}.} \bibinfo{year}{2012}\natexlab{b}.
\newblock \showarticletitle{Predicting common web application vulnerabilities
  from input validation and sanitization code patterns}. In
  \bibinfo{booktitle}{\emph{Proceedings of the 27th IEEE/ACM International
  Conference on Automated Software Engineering}}. \bibinfo{pages}{310--313}.
\newblock


\bibitem[\protect\citeauthoryear{Shar, Tan, and Briand}{Shar
  et~al\mbox{.}}{2013}]%
        {Shar:13}
\bibfield{author}{\bibinfo{person}{Lwin~Khin Shar}, \bibinfo{person}{Hee
  Beng~Kuan Tan}, {and} \bibinfo{person}{Lionel~C. Briand}.}
  \bibinfo{year}{2013}\natexlab{}.
\newblock \showarticletitle{Mining {SQL} Injection and Cross Site Scripting
  Vulnerabilities using Hybrid Program Analysis}. In
  \bibinfo{booktitle}{\emph{Proceedings of the 35th International Conference on
  Software Engineering}}. \bibinfo{pages}{642--651}.
\newblock


\bibitem[\protect\citeauthoryear{{Sink}}{{Sink}}{2017}]%
        {Sink:17}
\bibfield{author}{\bibinfo{person}{{Sink}}.} \bibinfo{year}{2017}\natexlab{}.
\newblock \bibinfo{title}{{XSS} Attacks: The Next Wave}.
\newblock
\newblock
\newblock
\shownote{https://snyk.io/blog/xss-attacks-the-next-wave/.}


\bibitem[\protect\citeauthoryear{Smith}{Smith}{2011}]%
        {Smith:11}
\bibfield{author}{\bibinfo{person}{Noah~A. Smith}.}
  \bibinfo{year}{2011}\natexlab{}.
\newblock \bibinfo{booktitle}{\emph{{Linguistic Structure Prediction}}}.
\newblock \bibinfo{publisher}{Graeme Hirst}.
\newblock


\bibitem[\protect\citeauthoryear{Son and Shmatikov}{Son and Shmatikov}{2011}]%
        {Son:11}
\bibfield{author}{\bibinfo{person}{Sooel Son} {and} \bibinfo{person}{Vitaly
  Shmatikov}.} \bibinfo{year}{2011}\natexlab{}.
\newblock \showarticletitle{{SAFERPHP}: Finding Semantic Vulnerabilities in
  {PHP} Applications}. In \bibinfo{booktitle}{\emph{Proceedings of the ACM
  SIGPLAN 6th Workshop on Programming Languages and Analysis for Security}}.
\newblock


\bibitem[\protect\citeauthoryear{{The Hacker News}}{{The Hacker News}}{2017a}]%
        {HackerNews:17a}
\bibfield{author}{\bibinfo{person}{{The Hacker News}}.}
  \bibinfo{year}{2017}\natexlab{a}.
\newblock \bibinfo{title}{It's 3 Billion! Yes, Every Single Yahoo Account Was
  Hacked In 2013 Data Breach}.
\newblock
\newblock
\newblock
\shownote{https://thehackernews.com/2017/10/yahoo-email-hacked.html.}


\bibitem[\protect\citeauthoryear{{The Hacker News}}{{The Hacker News}}{2017b}]%
        {HackerNews:17}
\bibfield{author}{\bibinfo{person}{{The Hacker News}}.}
  \bibinfo{year}{2017}\natexlab{b}.
\newblock \bibinfo{title}{WordPress Plugin Used by 300,000+ Sites Found
  Vulnerable to SQL Injection Attack}.
\newblock
\newblock
\newblock
\shownote{https://thehackernews.com/2017/06/wordpress-hacking-sql-injection.html.}


\bibitem[\protect\citeauthoryear{{threatpost}}{{threatpost}}{2017}]%
        {ThreatPost:17}
\bibfield{author}{\bibinfo{person}{{threatpost}}.}
  \bibinfo{year}{2017}\natexlab{}.
\newblock \bibinfo{title}{Million-Plus WordPress Sites Exposed by Vulnerable
  Plugin}.
\newblock
\newblock
\newblock
\shownote{https://threatpost.com/million-plus-wordpress-sites-exposed-by-vulnerable-plugin/123983/.}


\bibitem[\protect\citeauthoryear{Viterbi}{Viterbi}{1967}]%
        {Viterbi:67}
\bibfield{author}{\bibinfo{person}{A. Viterbi}.}
  \bibinfo{year}{1967}\natexlab{}.
\newblock \showarticletitle{Error Bounds for Convolutional Codes and an
  Asymptotically Optimum Decoding Algorithm}.
\newblock \bibinfo{journal}{\emph{IEEE Transactions on Information Theory}}
  \bibinfo{volume}{13}, \bibinfo{number}{2} (\bibinfo{date}{April}
  \bibinfo{year}{1967}), \bibinfo{pages}{260--269}.
\newblock


\bibitem[\protect\citeauthoryear{Walden, Doyle, Welch, and Whelan}{Walden
  et~al\mbox{.}}{2009}]%
        {Walden:09}
\bibfield{author}{\bibinfo{person}{James Walden}, \bibinfo{person}{Maureen
  Doyle}, \bibinfo{person}{Grant~A Welch}, {and} \bibinfo{person}{Michael
  Whelan}.} \bibinfo{year}{2009}\natexlab{}.
\newblock \showarticletitle{Security of open source web applications}. In
  \bibinfo{booktitle}{\emph{Proceedings of the 3rd International Symposium on
  Empirical Software Engineering and Measurement}}. \bibinfo{pages}{545--553}.
\newblock


\bibitem[\protect\citeauthoryear{Williams and Wichers}{Williams and
  Wichers}{2017}]%
        {owasp:17}
\bibfield{author}{\bibinfo{person}{J. Williams} {and} \bibinfo{person}{D.
  Wichers}.} \bibinfo{year}{2017}\natexlab{}.
\newblock \bibinfo{title}{{OWASP Top} 10 2017 -- The Ten Most Critical Web
  Application Security Risks}.
\newblock
\newblock


\bibitem[\protect\citeauthoryear{Witten, Frank, and Hall}{Witten
  et~al\mbox{.}}{2011}]%
        {Witten:11}
\bibfield{author}{\bibinfo{person}{Ian~H. Witten}, \bibinfo{person}{Eibe
  Frank}, {and} \bibinfo{person}{Mark~A. Hall}.}
  \bibinfo{year}{2011}\natexlab{}.
\newblock \bibinfo{booktitle}{\emph{Data Mining: Practical Machine Learning
  Tools and Techniques} (\bibinfo{edition}{3rd} ed.)}.
\newblock \bibinfo{publisher}{Morgan Kaufmann}.
\newblock


\bibitem[\protect\citeauthoryear{WordPress}{WordPress}{[n. d.]}]%
        {Wordpress:15}
\bibfield{author}{\bibinfo{person}{WordPress}.} \bibinfo{year}{[n.
  d.]}\natexlab{}.
\newblock
\newblock
\newblock
\shownote{{h}ttps://wordpress.org/.}


\bibitem[\protect\citeauthoryear{Yamaguchi, Golde, Arp, and Rieck}{Yamaguchi
  et~al\mbox{.}}{2014}]%
        {Yamaguchi:14}
\bibfield{author}{\bibinfo{person}{F. Yamaguchi}, \bibinfo{person}{N. Golde},
  \bibinfo{person}{D. Arp}, {and} \bibinfo{person}{K. Rieck}.}
  \bibinfo{year}{2014}\natexlab{}.
\newblock \showarticletitle{Modeling and Discovering Vulnerabilities with Code
  Property Graphs}. In \bibinfo{booktitle}{\emph{Proceedings of the 2014 IEEE
  Symposium on Security and Privacy}}. \bibinfo{pages}{590--604}.
\newblock


\bibitem[\protect\citeauthoryear{Yamaguchi, Maier, Gascon, and Rieck}{Yamaguchi
  et~al\mbox{.}}{2015}]%
        {Yamaguchi:15}
\bibfield{author}{\bibinfo{person}{F. Yamaguchi}, \bibinfo{person}{A. Maier},
  \bibinfo{person}{H. Gascon}, {and} \bibinfo{person}{K. Rieck}.}
  \bibinfo{year}{2015}\natexlab{}.
\newblock \showarticletitle{Automatic Inference of Search Patterns for
  Taint-Style Vulnerabilities}. In \bibinfo{booktitle}{\emph{Proceedings of the
  2015 IEEE Symposium on Security and Privacy}}. \bibinfo{pages}{797--812}.
\newblock


\bibitem[\protect\citeauthoryear{Yamaguchi, Wressnegger, Gascon, and
  Rieck}{Yamaguchi et~al\mbox{.}}{2013}]%
        {Yamaguchi:13}
\bibfield{author}{\bibinfo{person}{F. Yamaguchi}, \bibinfo{person}{C.
  Wressnegger}, \bibinfo{person}{H. Gascon}, {and} \bibinfo{person}{K. Rieck}.}
  \bibinfo{year}{2013}\natexlab{}.
\newblock \showarticletitle{Chucky: Exposing Missing Checks in Source Code for
  Vulnerability Discovery}. In \bibinfo{booktitle}{\emph{Proceedings of the
  20th ACM SIGSAC Conference on Computer Communications Security}}.
  \bibinfo{pages}{499--510}.
\newblock


\end{thebibliography}

\vspace{3mm}
\begin{description}
	\item \textbf{Ib\'{e}ria Medeiros} is an Assistant Professor in the Department of Informatics, at the Faculty of Sciences of University of Lisbon. She is a member of the Large-Scale Informatics Systems (LASIGE) Laboratory, and the Navigators research group. She holds a PhD in Computer Science by the Faculty of Sciences of University of Lisbon. Currently, she is the principal investigator of the SEAL national project, has been participating in DiSIEM European project and REDBOOK national project, and participate in SEGRID European project. She is author of tools for software security, which WAP (Web Application Protection) is the most known and an OWASP project. Her research interests are concerned with software security, source code static analysis, vulnerability detection, data mining and machine learning, and security. More information about her at http://www.di.fc.ul.pt/$\sim$imedeiros/.
	
	\vspace{3mm}
	\item \textbf{Nuno Neves} is Professor at the Department of Computer Science, Faculty of Sciences of the University of Lisboa. He leads the Navigators research group and he is on the scientific board of the LASIGE research unit. His main research interests are in security and dependability aspects of distributed systems. Currently, he is investigator in several national and EU projects, such as SEAL and uPVN. His work has been recognized in several occasions, for example with the IBM Scientific Prize and the William C. Carter award. He is on the editorial board of the International Journal of Critical Computer-Based Systems. More information about him can be found at http://www.di.fc.ul.pt/$\sim$nuno/.
	
	\vspace{3mm}
	\item \textbf{Miguel Correia} is an Associate Professor with Habilitation at Instituto Superior T\'{e}cnico (IST) of Universidade de Lisboa (ULisboa), and a Senior Researcher at INESC-ID in the Distributed Systems Group (GSD). He has been involved in several international and national research projects related to cybersecurity, including the SPARTA, QualiChain, SafeCloud, PCAS, TCLOUDS, ReSIST, CRUTIAL, and MAFTIA European projects. He has more than 150 publications and is Senior Member of the IEEE. More information about him at http://www.gsd.inesc-id.pt/$\sim$mpc/.
\end{description}

\end{document}